\RequirePackage{etoolbox}
\csdef{input@path}{{style/}{graphics/}}
\documentclass[aap,MSNbibl,dvips]{arximspdf}

\doi{10.1214/14-AAP1005} 
\volume{25}
\issue{2}
\pubyear{2015}
\firstpage{600}
\lastpage{631}

\makeatletter

\newcommand{\eqref}[1]{(\ref{#1})}
\newtheorem{lemma}{Lemma}[section]
\newproclaim{definition}{Definition}[section]
\newtheorem{theorem}{Theorem}[section]
\newtheorem{proposition}{Proposition}[section]
\newproclaim{remark}{Remark}[section]

\newproclaim{assumption}{Assumption}
\newproclaim{example}{Example}
\newtheorem{corollary}{Corollary}[section]

\newcommand{\E}{\mathbb{E}}
\makeatother

\begin{document}
\begin{frontmatter}

\title{Limit theorems for nearly unstable Hawkes~processes}
\runtitle{Limit theorems for nearly unstable Hawkes processes}
\begin{aug}
\author[A]{\fnms{Thibault} \snm{Jaisson}\corref{}\ead[label=e1]{thibault.jaisson@polytechnique.edu}}
\and
\author[B]{\fnms{Mathieu} \snm{Rosenbaum}\ead[label=e2]{mathieu.rosenbaum@upmc.fr}}
\runauthor{T. Jaisson and M. Rosenbaum}
\affiliation{\'Ecole Polytechnique Paris and Universit\'e Pierre et
Marie Curie (Paris 6)}
\address[A]{Centre de Math\'{e}matiques Appliqu\'{e}es\\
CNRS UMR 7641\\
Ecole Polytechnique Paris\\
Route de Saclay\\
91128 Palaiseau Cedex\\
France\\
\printead{e1}} 
\address[B]{Laboratoire de Probabilit\'{e}s\\
\quad et Mod\`{e}les Al\'{e}atoires\\
CNRS UMR 7599\\
Universit\'{e} Pierre et Marie Curie\\
4 Place Jussieu\\
75 252 Paris Cedex\\
France\\
\printead{e2}}
\end{aug}

\received{\smonth{10} \syear{2013}}

%
\begin{abstract}
Because of their tractability and their natural interpretations in term
of market quantities, Hawkes processes are nowadays widely used in
high-frequency finance. However, in practice, the statistical
estimation results seem to show that very often, only \textit{nearly
unstable Hawkes processes} are able to fit the data properly. By nearly
unstable, we mean that the $L^1$ norm of their kernel is close to
unity. We study in this work such processes for which the stability
condition is almost violated. Our main result states that after
suitable rescaling, they asymptotically behave like integrated
Cox--Ingersoll--Ross models. Thus, modeling financial order flows as
nearly unstable Hawkes processes may be a good way to reproduce both
their high and low frequency stylized facts. We then extend this result
to the Hawkes-based price model introduced by Bacry et al. [\emph{Quant. Finance}
\textbf{13} (2013) 65--77]. We show that under a similar
criticality condition, this process converges to a Heston model. Again,
we recover well-known stylized facts of prices, both at the
microstructure level and at the macroscopic scale.
\end{abstract}

%

\end{frontmatter}

\section{Introduction}\label{intro}

A Hawkes process $(N_t)_{t\geq0}$ is a self exciting point process,
whose intensity at time $t$, denoted by $\lambda_t$, is of the form
\[
\lambda_t=\mu+\sum_{0<J_i<t}
\phi(t-J_i)=\mu+\int_{(0,t)}\phi(t-s)\,dN_s,
\]
where $\mu$ is a positive real number, $\phi$ a regression kernel and
the $J_i$ are the points of the process before time $t$; see Section~\ref{s2}
for more accurate definitions.
These processes were introduced in 1971 by Hawkes (see \cite
{hawkes1971point,hawkes1971spectra,hawkes1974cluster}) for the purpose
of modeling earthquakes and their aftershocks; see \cite
{adamopoulos1976cluster}. However, they are also used in various other
disciplines. In particular, in recent years, with the availability of
(ultra) high-frequency data, finance has become one of the main domains
of application of Hawkes processes.

The introduction of Hawkes processes in finance is probably
due to Chavez-Demoulin et al. (see \cite{chavez2005point}), in the
context of value at risk estimation, and to Bowsher (see \cite
{bowsher2007modelling}), who jointly studied transaction times and
midquote changes, using the Hawkes framework. Then, in \cite
{bauwens2004dynamic}, Bauwens and Hautsch built so-called latent factor
intensity Hawkes models and applied them to transaction data. Another
pioneer of this type of approach is Hewlett. He considered in \cite
{hewlett2006clustering} the particular case of the foreign exchange
rates market for which he fitted a bivariate Hawkes process on buy and
sell transaction data. More recently, Bacry et al. have developed a
microstructure model for midquote prices based on the difference of two
Hawkes processes; see \cite{bacry2013modelling}. Moreover, Bacry and
Muzy have extended this approach in \cite{bacry2013hawkes} where they
design a framework enabling to study market impact. Beyond midquotes
and transaction prices, full limit order book data (not only market
orders but also limit orders and cancellations) have also been
investigated through the lenses of Hawkes processes. In particular,
Large uses in \cite{large2007measuring} a ten-variate multidimensional
Hawkes process to this purpose. Note that besides microstruture
problems, Hawkes processes have also been introduced in the study of
other financial issues such as daily data analysis (see \cite
{embrechts2011multivariate}), financial contagion (see \cite
{ait2010modeling}) or credit risk; see \cite{errais2010affine}.

Hawkes processes have become popular in financial modeling
for two main reasons. First, these processes represent a very natural
and tractable extension of Poisson processes. In fact, comparing point
processes and conventional time series, Poisson processes are often
viewed as the counterpart of i.i.d. random variables, whereas Hawkes
processes play the role of autoregressive processes; see \cite
{daley2007introduction} for more details about this analogy. Another
explanation for the appeal of Hawkes processes is that
it is often easy to give a convincing interpretation to such modeling.
To do so, the branching structure of Hawkes processes is quite helpful.
Recall that under the assumption $\|\phi\|_{1}<1$, where $\|\phi\|_{1}$
denotes the $L^1$ norm of $\phi$,
Hawkes processes can be represented as a population process where
migrants arrive according to a Poisson process with parameter $\mu$.
Then each migrant gives birth to children according to a nonhomogeneous
Poisson process with intensity function~$\phi$, these children also
giving birth to children according to the same nonhomogeneous Poisson
process; see \cite{hawkes1974cluster}. Now consider, for example, the
classical case of buy (or sell) market orders, as studied in several of
the papers mentioned above. Then migrants can be seen
as exogenous orders whereas children are viewed as orders triggered by
other orders.

Beyond enabling us to build this population dynamics
interpretation, the assumption $\|\phi\|_{1}<1$ is crucial in the study
of Hawkes processes. To fix ideas, let us place ourselves in the
classical framework where the Hawkes process $(N_t)$ starts at
$-\infty$. In that case, if one wants to get a stationary intensity
with finite first moment, then the condition $\|\phi\|_{1}<1$ is
necessary. Furthermore, even in the nonstationary setting, this
condition is usually required in order to obtain classical ergodic
properties for the process; see \cite{bacry2012scaling}. For these
reasons, this condition is often called a stability condition in the
Hawkes literature.

From a practical point of view, a lot of interest has been
recently devoted to the parameter $\|\phi\|_{1}$. For example,
Hardiman, Bercot and Bouchaud (see \cite{hardiman2013critical}) and
Filimonov and Sornette (see \cite
{filimonov2012quantifying,filimonov2013apparent}), use the branching
interpretation of Hawkes processes on midquote data in order to measure
the so-called degree of endogeneity of the market. This degree is
simply defined by $\|\phi\|_{1}$, which is also called branching ratio.
The intuition behind this interpretation of $\|\phi\|_{1}$ goes as
follows: The parameter
$\|\phi\|_{1}$ corresponds to the average number of children of an
individual, $\|\phi\|_{1}^2$ to the average number of grandchildren of
an individual$,\ldots.$ Therefore, if we call cluster the descendants of a
migrant, then the average size of a cluster is given by $\sum_{k\geq
1}\|\phi\|_{1}^k=\|\phi\|_{1}/(1-\|\phi\|_{1})$. Thus, in the financial
interpretation, the average proportion of endogenously triggered events
is $\|\phi\|_{1}/(1-\|\phi\|_{1})$ divided by $1+\|\phi\|_{1}/(1-\|\phi
\|_{1})$, which is equal to $\|\phi\|_{1}$.

This branching ratio can be measured using parametric and
nonparametric estimation methods for Hawkes processes; see \cite
{ogata1978asymptotic,ogata1983likelihood} for likelihood based methods
and \cite{bacry2012non,reynaud2010adaptive} for functional estimators
of the function $\phi$. In \cite{hardiman2013critical}, very stable
estimations of $\|\phi\|_{1}$ are reported for the E mini S\&P futures
between 1998 and 2012, the results being systematically close to one.
In \cite{filimonov2012quantifying}, values of order 0.7--0.8 are
obtained on several assets. A debate on the validity of these results
is currently ongoing between the two groups. In particular, it is
argued in \cite{hardiman2013critical} that the choice of exponential
kernels in \cite{filimonov2012quantifying} may lead to spurious
results, whereas various bias that could affect the study in \cite
{hardiman2013critical} are underlined in \cite{filimonov2013apparent}.
In any case, we can remark that both groups find values close to one
for $\|\phi\|_{1}$, which is consistent with the results of \cite
{bacry2012non}, where estimations are performed on Bund and Dax
futures.

This seemingly persistent statistical result should
definitely worry users of Hawkes processes. Indeed, it is rarely
suitable to apply a statistical model where the parameters are pushed
to their limits. In fact, these obtained values for $\|\phi\|_{1}$ on
empirical data are not really surprising. Indeed, one of the
best-documented stylized facts in high-frequency finance is the
persistence (or long memory) in flows and market activity measures;
see, for example, \cite{bouchaud2004fluctuations,lillo2004long}. Usual
Hawkes processes, in the same way as autoregressive processes, can only
exhibit short-range dependence, failing to reproduce this classical
empirical feature; see \cite{jaisson2013impact} for details.

In spite of their relative inadequacy with market data,
Hawkes processes possess so many appealing properties that one could
still try to apply them in some specific situations. In \cite
{hardiman2013critical}, it is suggested to use the ``without
ancestors'' version of Hawkes processes introduced by Br\'emaud and
Massouli\'e in \cite{bremaud2001hawkes}. For such processes, $\|\phi\|
_{1}=1$, but in order to preserve stationarity and a finite expectation
for the intensity, one needs to have $\mu=0$.
This is probably a relevant approach. However setting the parameter $\mu
$ to $0$ is not completely satisfying since this parameter has a nice
interpretation (exogenous orders). Moreover it is not found to be equal
to zero in practice, see \cite{hardiman2013critical}. Finally, a
time-varying $\mu$ is an easy way to reproduce seasonalities observed
on the market; see \cite{bacry2013hawkes} (however, for simplicity, we
work in this paper with a constant $\mu>0$).


These empirical measures of $\|\phi\|_{1}$, close to one, are
the starting point of this work. Indeed, our aim is to study the
behavior at large time scales of nearly unstable Hawkes processes,
which correspond to these estimations. More precisely, we consider a
sequence of Hawkes processes observed on $[0,T]$, where $T$ goes to
infinity. In the case of a fixed kernel (not depending on $T$) with
norm strictly smaller than one, scaling limits of Hawkes processes have
been investigated in \cite{bacry2012scaling}, see also
\cite{zhu2013central} for the case of non linear Hawkes processes.
In this framework, Bacry
et al. obtain a deterministic limit for the properly normalized
sequence of Hawkes processes, as it is the case for suitably rescaled
Poisson processes. In their price model consisting in the difference of
two Hawkes processes, a Brownian motion (with some volatility) is found
at the limit. These two results are in fact quite intuitive. Indeed, in
the same way as Poisson processes and autoregressive models, Hawkes
processes enjoy short memory properties. In this work, we show that
when the Hawkes processes are nearly unstable, these weakly
dependent-like behaviors are no longer observed at intermediate time
scales. To do so, we consider that the kernels of the Hawkes processes
depend on $T$. More precisely, we translate the near instability
condition into the assumption that the norm of the kernels tends to one
as the observation scale $T$ goes to infinity.

Our main theorem states that when the norm of the kernel
tends to one at the right speed (meaning that the observation scale and
kernel's norm balance in a suitable way), the limit of our sequence of
Hawkes processes is no longer a deterministic process, but an
integrated Cox--Ingersoll--Ross process (CIR for short), as introduced
in \cite{cox1985theory}. In practice, it means that when observing a
Hawkes process with kernel's norm close to one at appropriate time
scale, it looks like an integrated CIR. Furthermore, for the price
model defined in \cite{bacry2013modelling}, in the limit, the Brownian
motion obtained in \cite{bacry2012scaling} is replaced by a Heston
model; see \cite{heston1993closed} for definition. This is probably
more in agreement with empirical data.

The paper is organized as follows. The assumptions and main
results, notably the convergence toward an integrated CIR are given
in Section~\ref{s2}. The case of the difference of two Hawkes processes
is studied in Section~\ref{s3}. The proofs are relegated to Section~\ref{proofs}.

\section{Scaling limits of nearly unstable Hawkes processes}\label{s2}

We give in this section our main results about the limiting behavior of
a sequence of nearly unstable Hawkes processes.
We start by presenting our assumptions and defining our asymptotic setting.

\subsection{Assumptions and asymptotic framework}

We consider a sequence of point processes $(N_t^T)_{t\geq0}$ indexed
by $T$.\footnote{Of course by $T$ we implicitly means $T_n$ with $n\in
\mathbb N$ tending to infinity.} For a given $T$, $(N_t^T)$ satisfies
$N_0^T=0$, and the process is observed on the time interval $[0,T]$.
Furthermore, our asymptotic setting is that the observation scale $T$
goes to infinity. The intensity process $(\lambda_t^T)$ is defined for
$t\geq0$ by
\[
\lambda_t^T=\mu+\int_0^t
\phi^T(t-s) \,dN^T_s,
\]
where $\mu$ is a positive real number and $\phi^T$ a nonnegative
measurable function on $\mathbb{R}^+$ which satisfies $\|\phi^T\|
_{1}<+\infty$. For a given $T$, the process $(N_t^T)$ is defined on a
probability space $(\Omega^T,\mathcal{F}^T,\mathbb{P}^T)$ equipped with
the filtration $(\mathcal{F}_t^T)_{t\in[0,T]}$, where $\mathcal{F}_t^T$
is the $\sigma$-algebra generated by $(N_s^T)_{s\leq t}$. Moreover we
assume that for any $0\leq a<b\leq T$ and $A\in\mathcal{F}_a^T$,
\[
\E\bigl[\bigl(N_b^T-N^T_{a}\bigr)
\mathrm{1}_A\bigr]=\E\biggl[\int_a^b
\lambda^T_s\mathrm{1}_A \,ds\biggr],
\]
which sets $\lambda^T$ as the intensity of $N^T$. In particular,
if we denote by $(J_n^T)_{n\geq1}$ the jump times of $(N_t^T)$, the process
\[
N^T_{t\wedge J_n^T}-\int_0^{t\wedge J_n^T}
\lambda_s^T\,ds
\]
is a martingale and the law of $N^T$ is characterized by $\lambda^T$.
From Jacod \cite{jacod1975multivariate}, such construction can be done.
The process $N^T$ is called a Hawkes process.

Let us now give more specific assumptions on the function
$\phi^T$. We denote by $\|.\|_{\infty}$ the $L^{\infty}$ norm on
$\mathbb{R}^+$.

\begin{assumption}\label{h1}
For $t\in\mathbb{R}^+$,
\[
\phi^T(t)=a_T\phi(t),
\]
where $(a_T)_{T\geq0}$ is a sequence of positive numbers converging to
one such that for all $T$, $a_T<1$ and $\phi$ is a nonnegative
measurable function such that
\[
\int_0^{+\infty}\phi(s) \,ds=1\quad\mbox{and}\quad  \int
_0^{+\infty}s\phi(s) \,ds=m<\infty.
\]
Moreover, $\phi$ is differentiable with derivative $\phi'$ such that $\|
\phi'\|_{\infty}<+\infty$ and $\|\phi'\|_{1}<+\infty$.
\end{assumption}

\begin{remark}
Note that under Assumption~\ref{h1}, $\|\phi\|_{\infty}$ is finite.
\end{remark}

Thus, the form of the function $\phi^T$ depends on $T$ so
that its shape is fixed, but its $L^1$ norm varies with $T$. For a
given $T$, this $L^1$ norm is equal to $a_T$ and so is smaller than
one, implying that the stability condition is in force. Note that in
this framework, we have almost surely no explosion,\footnote{In fact,
for a Hawkes process, the no explosion property can be obtained under
weaker conditions, for example, $\int_0^t\phi(s)\,ds<\infty$ for any
$t>0$; see \cite{bacry2012scaling}.}
\[
\label{explosion}\mathop{\operatorname {lim}}_{n\rightarrow+\infty}J_n^T=+
\infty.
\]
However, remark that we do not work in the stationary setting since our
process starts at time $t=0$ and not at $t=-\infty$.

The case where $\|\phi^T\|_1$ is larger than one corresponds
to the situation where the stability condition is violated. Since
$a_T=\|\phi^T\|_1<1$ tends to one, our framework is a way to get close
to instability. Therefore we call our processes nearly unstable Hawkes
processes. There are of course many other ways to make the $L^1$ norm
of $\phi^T$ converge to one than the multiplicative manner used here.
However, this parametrization is sufficient for applications and very
convenient to illustrate the different regimes that can be obtained.

\subsection{Observation scales}

In our framework, two parameters degenerate at infinity: $T$ and
$(1-a_T)^{-1}$. The relationship between these two sequences will
determine the scaling behavior of the sequence of Hawkes processes.
Recall that it is shown in \cite{bacry2012scaling} that when $\|\phi\|
_{1}$ is fixed and smaller than one, after appropriate scaling, the
limit of the sequence of Hawkes processes is deterministic, as it is,
for example, the case for Poisson processes. In our setting, if $1-a_T$
tends ``slowly'' to zero, we can expect the same result. Indeed, we may
have $T$ large enough so that we reach the asymptotic regime and for
such $T$, $a_T$ is still sufficiently far from unity. This is precisely
what happens, as stated in the next theorem.

\begin{theorem}
\label{Th1}
Assume $T(1-a_T)\rightarrow+\infty$. Then, under Assumption~\ref{h1},
the sequence of Hawkes processes is asymptotically deterministic, in
the sense that the following convergence in $L^2$ holds:
\[
\mathop{\operatorname{sup}}_{v\in[0,1]}\frac{1-a_T}{T}\bigl|N^T_{Tv}-
\E \bigl[N^T_{Tv}\bigr]\bigr|\rightarrow0.
\]
\end{theorem}

On the contrary, if $1-a_T$ tends too rapidly to zero, the
situation is likely to be quite intricate. Indeed, for given $T$, the
Hawkes process may already be very close to instability whereas $T$ is
not large enough to reach the asymptotic regime. The last case, which
is probably the most interesting one, is the intermediate case, where
$1-a_T$ tends to zero in such a manner that a nondeterministic scaling
limit is obtained, while not being in the preceding degenerate setting.
We largely detail this situation in the next subsection.

\subsection{Nondegenerate scaling limit for nearly unstable Hawkes processes}

We give in this section our main result: a nondegenerate scaling limit
for a sequence of properly renormalized nearly unstable Hawkes
processes. Before giving this theorem, we wish to provide intuitions on
how it is derived. Let $M^T$ be the martingale process associated to
$N^T$, that is, for $t\geq0$,
\[
M^T_t=N^T_t-\int
_0^t\lambda^T_s\,ds.
\]
We also set $\psi^T$ the function defined on $\mathbb{R}^+$ by
\[
\psi^T(t)=\sum_{k=1}^\infty\bigl(
\phi^T\bigr)^{*k}(t),
\]
where $(\phi^T)^{*1}=\phi^T$ and for $k\geq2$, $(\phi^T)^{*k}$ denotes
the convolution product of $(\phi^T)^{*(k-1)}$ with the function $\phi
^T$. Note that $\psi^T(t)$ is well defined since $\|\phi^T\|_1<1$. In
the sequel, it will be convenient to work with another form for the
intensity. We have the following result, whose proof is given in
Section~\ref{proofs}.

\begin{proposition}
\label{formlambdamart}
For all $t\geq0$, we have
\[
\lambda_t^T=\mu+\int_0^t
\psi^T(t-s) \mu \,ds+\int_0^t
\psi^T(t-s)\,dM^T_s.
\]
\end{proposition}

Now recall that we observe the process $(N_t^T)$ on $[0,T]$.
In order to be able to give a proper limit theorem, where the processes
live on the same time interval, we rescale our processes so that they
are defined on $[0,1]$. To do so, we consider for $t\in[0,1]$
\[
\lambda_{tT}^T=\mu+\int_0^{tT}
\psi^T(Tt-s) \mu \,ds+\int_0^{tT} \psi
^T(Tt-s)\,dM^T_s.
\]

For the scaling in space, a natural multiplicative factor is
$(1-a_T)$. Indeed, in the stationary case, the expectation of $\lambda
_t^T$ is $\mu/(1-\|\phi^T\|_1)$. Thus the order of magnitude of the
intensity is $(1-a_T)^{-1}$. This is why we define
%
\begin{equation}
\label{intens} C^T_t=\lambda_{tT}^T(1-a_T).
\end{equation}

Understanding the asymptotic behavior of $C^T_t$ will be the
key to the derivation of a suitable scaling limit for our sequence of
renormalized processes. We will see that this behavior is closely
connected to that of the function $\psi^T$. About $\psi^T$, one can
first remark that the function defined for $x\geq0$ by
%
\begin{equation}
\label{rho} \rho^T(x)=T\frac{\psi^T}{\|\psi^T\|_1}(Tx)
\end{equation}
is the density of the random variable
\[
X^T=\frac{1}{T} \sum_{i=1}^{I^T}
X_i,
\]
where the $(X_i)$ are i.i.d. random variables with density $\phi$ and
$I^T$ is a geometric random variable with parameter $1-a_T$ ($\forall
k>0, \mathbb{P}[I^T=k]=(1-a_T)(a_T)^{k-1}$). Now let $z\in\mathbb{R}$.
The characteristic function of the random variable $X^T$, denoted by~$\widehat{\rho}^T$, satisfies
\begin{eqnarray*}
\widehat{\rho}^T(z)&=&\E\bigl[e^{izX^T}\bigr]=\sum
_{k=1}^\infty(1-a_T) (a_T)^{k-1}
\E\bigl[e^{i({z}/{T})\sum_{i=1}^kX_i}\bigr]
\\
&=&\sum_{k=1}^\infty(1-a_T)
(a_T)^{k-1} \biggl(\hat{\phi}\biggl(\frac
{z}{T}\biggr)
\biggr)^{k}=\frac{\hat{\phi}({z}/{T})}{1-({a_T}/{(1-a_T)})(\hat
{\phi}({z}/{T})-1)},
\end{eqnarray*}
where $\hat{\phi}$ denotes the characteristic function of $X_1$. Since
\[
\int_0^{+\infty}s\phi(s) \,ds=m<\infty,
\]
the function $\hat{\phi}$ is continuously differentiable with first
derivative at point zero equal to $im$. Therefore, using that $a_T$ and
$\hat{\phi}(\frac{z}{T})$ both tend to one as $T$ goes to infinity,
$\widehat{\rho}^T(z)$ is equivalent to
\[
\frac{1}{1-{izm}/{(T(1-a_T))}}.
\]

Thus, we precisely see here that the suitable regime so that
we get a nontrivial limiting law for $X^T$ is that there exists $\lambda
>0$ such that
%
\begin{equation}
\label{regime} T(1-a_T)\mathop{\rightarrow}_{T\rightarrow+\infty}\lambda.
\end{equation}

When \eqref{regime} holds, we write $d_0=m/\lambda$. In fact
we have just proved the following result.

\begin{proposition}
\label{renyi}
Assume that \eqref{regime} holds. Under Assumption~\ref{h1}, the
sequence of random variable $X^T$
converges in law toward an exponential random variable with parameter $1/d_0$.
\end{proposition}

This simple result is of course not new. For example these
types of geometric sums of random variables are studied in detail in
\cite{kalashnikov1997geometric}. Note also that when $X_1$ is
exponentially distributed, $X^T$ is also exponentially distributed,
even for a fixed $T$.

Assume from now on that \eqref{regime} holds and set
$u_T=T(1-a_T)/\lambda$ (so that $u_T$ goes to one). Proposition~\ref
{renyi} is particularly important since it gives us the asymptotic
behavior of $\psi^T$ in this setting. Indeed, it tells us that
\[
\psi^T(Tx)=\rho^T(x) \frac{a_T}{\lambda u_T} \approx
\frac{\lambda
}{m}e^{-x({\lambda}/{m})}\frac{1}{\lambda}=\frac{1}{m}e^{-x
({\lambda}/{m})}.
\]
Let us now come back to the process $C_t^T$, which can be written
%
\begin{equation}
\label{C2} C_t^T=(1-a_T) \mu+\mu\int
_0^{t} u_T\lambda\psi^T
(Ts) \,ds+\int_0^{t} \sqrt{\lambda}
\psi^T\bigl(T(t-s)\bigr) \sqrt{C_s^T}\,dB^T_s,\hspace*{-16pt}
\end{equation}
with
%
\begin{equation}
\label{defB}B_t^T=\frac{1}{\sqrt{T}}
\sqrt{u_T}\int_0^{tT}
\frac{dM_s^T}{\sqrt{\lambda^T_s}}.
\end{equation}
By studying its quadratic variation, we will show that $B^T$ represents
a sequence of martingales which converges to a Brownian motion.
So, heuristically replacing $B^T$ by a Brownian motion $B$ and $\psi
^T(Tx)$ by $\frac{1}{m}e^{-x{{\lambda}/{m}}}$ in \eqref{C2}, we get
\[
C_t^{\infty}=\mu\bigl(1-e^{-t({\lambda}/{m})}\bigr)+
\frac{\sqrt{\lambda}}{m}\int_0^t e^{-(t-s)({\lambda}/{m})}
\sqrt{C_s^{\infty}} \,dB_s.
\]

Applying It\^o's formula, this gives
\[
C_t^{\infty}=\int_0^t
\bigl(\mu-C_s^{\infty}\bigr)\frac{\lambda}{m} \,ds +
\frac
{\sqrt{\lambda}}{m}\int_0^t \sqrt
{C_s^{\infty}} \,dB_s,
\]
which precisely corresponds to the stochastic differential equation
(SDE) satisfied by a CIR process.

Before stating the theorem which makes the preceding
heuristic derivation rigorous, we consider an additional assumption.

\begin{assumption}
\label{h2}
There exists $K_\rho>0$ such that for all $x\geq0$ and $T>0$,
\[
\bigl|\rho^T(x)\bigr|\leq K_\rho.
\]
\end{assumption}

Note that Assumption~\ref{h2} is in fact not really
restrictive. Indeed, if $\phi$ is decreasing, then any $\rho^T$ is
decreasing. Thus, since $|\rho^T(0)|$ is bounded, Assumption~\ref{h2}
holds in this case. Also, from \cite{petrov1975sums} (page 214, point
5), we get that if $\|\phi\|_{\infty}<\infty$ and $\int_0^{+\infty}
|s|^3\phi(s) \,ds<+\infty$, then Assumption~\ref{h2} follows. From \cite
{kalashnikov1997geometric} (Chapter~5, Lemma~4.1), another sufficient
condition to get Assumption~\ref{h2} is that the random variable $X_1$
with density $\phi$ can be written (in law) under the form $X_1=E+Y$,
where $E$ follows an exponential law with parameter $\gamma>0$ and $Y$
is independent of $E$. We now give our main theorem.

\begin{theorem}\label{main}
Assume that \eqref{regime} holds. Under Assumptions \ref{h1} and \ref{h2},
the sequence of renormalized Hawkes intensities $(C^T_t)$ defined in
\eqref{intens} converges in law, for the Skorohod topology, toward the
law of the unique strong solution of the following Cox--Ingersoll--Ross
stochastic differential equation on $[0,1]$:
\[
X_t=\int_0^t (
\mu-X_s)\frac{\lambda}{m} \,ds +\frac{\sqrt{\lambda}}{m}\int
_0^t \sqrt{X_s} \,dB_s.
\]
Furthermore, the sequence of renormalized Hawkes process
\[
V_t^T=\frac{1-a_T}{T}N^T_{tT}
\]
converges in law, for the Skorohod topology, toward the process
\[
\int_0^tX_s \,ds,\qquad t\in[0,1].
\]
\end{theorem}

\subsection{Discussion}

\begin{itemize}
\item Theorem~\ref{main} implies that when $\|\phi\|_1$ is close to
$1$, if the observation time $T$ is suitably chosen [that is of order
$1/(1-\|\phi\|_1)$], a nondegenerate behavior (neither explosive, nor
deterministic) can be obtained for a rescaled Hawkes process.

\item This can, for example, be useful for the statistical estimation
of the parameters of a Hawkes process. Indeed, designing an estimating
procedure based on the fine scale properties of a Hawkes process is a
very hard task: Nonparametric methods are difficult to use and present
various instabilities (see \cite{bacry2012non,filimonov2013apparent}),
whereas parametric approaches are of course very sensitive to model
specifications; see \cite{filimonov2013apparent,hardiman2013critical}.
Considering an intermediate scale, where the process behaves like a CIR
model, one can use statistical methods specifically developed in order
to estimate CIR parameters; see \cite{alaya2012parameter} for a survey.
Of course, only the parameters $\lambda$, $m$ and $\mu$ can be
recovered this way. Therefore, there is clearly an information loss in
this approach. However, it still gives us access to quantities which
are important in practice; see Section~\ref{intro}. In some sense, it
can be compared to the extreme value theory based method for extreme
quantile estimation, where one assumes that the random variables of an
i.i.d. sample belong to some max stable attraction domain. Indeed,
these two methods lie between a fully parametric one, where a
parametric form is assumed (for the law of the random variables or the
function $\phi$), and a fully nonparametric one, where a functional
estimator (of the repartition function or of $\phi$) is used in order
to reach the quantity of interest (the quantile or the $L^1$ norm of
$\phi$).

\item CIR processes are a very classical way to model stochastic
(squared) volatilities in finance; see the celebrated Heston model \cite
{heston1993closed}. Also, it is widely acknowledged that there exists a
linear relationship between the cumulated order flow and the integrated
squared volatility; see, for example, \cite{wyart2008relation}.
Therefore, our setting where $\|\phi\|_1$ is close to one and the
limiting behavior obtained in Theorem~\ref{main} seem in good agreement
with market data.

\item For the stationary version of a Hawkes process, one can show that
the variance of $N_T^T$ is of order $T(1-\|\phi^T\|_1)^{-3}$; see, for
example, \cite{bremaud2001hawkes}. Therefore, if $T(1-a_T)$ tends to
zero, that is, $\|\phi^T\|_1$ goes rapidly to one, then the variance of
$\frac{(1-a_T)}{T}N_T^T$ blows up as $T$ goes to infinity. This
situation is therefore very different from the one studied here and is
therefore out of the scope of this paper.

\item The assumption $\int_0^{+\infty} s\phi(s)\,ds<+\infty$ is crucial
in order to approximate $\psi^T$ by an exponential function using
Proposition~\ref{renyi}. Let us now consider the fat tail case where
the preceding integral is infinite. More precisely, let us take a
function $\phi$ which is of order $\frac{1}{x^{1+\alpha}}$, $0<\alpha
<1$, as $x$ goes to infinity. In this case, following the proof of
Proposition~\ref{renyi}, we can show the following result, where we
borrow the notation of Proposition~\ref{renyi}.

\begin{proposition}\label{LM}
Let $E^\alpha_C$ be a random variable whose characteristic function satisfies
\[
\E\bigl[e^{izE^\alpha_C}\bigr]=\frac{1}{1-C(iz)^\alpha}.
\]
Assume $\hat{\phi}(z)-1\sim_{0} \sigma(iz)^\alpha$ for some $\sigma
>0$, $0<\alpha<1$ and $(1-a_T)T^\alpha\rightarrow\lambda>0$. Then
$X^T$ converges in law toward the random variable $E^\alpha_{{\sigma
}{\lambda}}$.
\end{proposition}

Thus, when the shape of the kernel is of order $x^{-(1+\alpha
)}$, the ``right'' observation scale is no longer $T\sim1/(1-\|\phi\|
_1)$, but $T\sim1/(1-\|\phi\|_1)^{{1}/{\alpha}}$. Remark also that
if we denote by $E_{\alpha,\beta}$ the $(\alpha,\beta)$ Mittag--Leffler
function, that is,
\[
E_{\alpha,\beta}(z)=\sum_{n=1}^{\infty}
\frac{z^n}{\Gamma(\alpha n+\beta)}
\]
(see, e.g., \cite{shah2013results}), then the density $\phi
^\alpha_C$ of $E^\alpha_C$ is linked to this function since
\[
\phi^\alpha_1(x)=x^{\alpha-1}E_{\alpha,\alpha}
\bigl(-x^\alpha\bigr).
\]

Now let us consider the asymptotic setting where $\mu^T=\mu
T^{\alpha-1}$, $\phi^T=a_T\phi$ with $a_T=1-\frac{\lambda}{T^\alpha}$
and $\phi$ as in Proposition~\ref{LM}. If we apply the same heuristic
arguments as those used in Section~\ref{s2} to the renormalized
intensity
\[
C^T_t=\frac{\lambda^T_{tT}(1-a_T)}{T^{\alpha-1}},
\]
we get the following type of limiting law for our sequence of Hawkes
intensities:
\[
X_t=\mu\int_0^t
\phi^\alpha_{{\sigma} /{\lambda}}(t-s)\,ds+\int
_0^t\phi ^\alpha_{{\sigma}/
{\lambda}}(t-s)\frac{1}{\sqrt{\lambda}}\sqrt{X_s}\,dB_s.
\]

These heuristic arguments are, however, far from a proof. Indeed, in
this case, we probably have to deal with a non semi-martingale limit.
Furthermore, tightness properties which are important in the proofs of
this paper are much harder to show (in particular the function $\phi
^\alpha_{C}$ is not bounded). We leave this case for further research.

\item In the classical time series setting, let us mention the paper
\cite{barczy2011asymptotic} where the authors study the asymptotic
behavior of unstable integer-valued autoregressive model (INAR
processes). In this case, CIR processes also appear in the limit. This
is, in fact, not so surprising since INAR processes share some
similarities with Hawkes processes. In particular, they can somehow be
viewed as Hawkes processes for which the kernel would be a sum of Dirac
functions.
\end{itemize}

\section{Extension of Theorem \texorpdfstring{\protect\ref{main}}{2.2} to a price model}\label{s3}

In the previous section, we have studied one-dimensional nearly
unstable Hawkes processes. For financial applications, they can, for
example, be used to model the arrival of orders when the number of
endogenous orders is much larger than the number of exogenous orders,
which seems to be the case in practice; see \cite
{filimonov2012quantifying,hardiman2013critical}. In this section, we
consider the high-frequency price model introduced in \cite
{bacry2013modelling}, which is essentially defined as a difference of
two Hawkes processes. Using the same approach as in Theorem~\ref{main},
we investigate the limiting behavior of this model when the stability
condition is close to saturation.

\subsection{A Hawkes-based price model}

In \cite{bacry2013modelling}, tick-by-tick moves of the midprice
$(P_t)_{t\geq0}$ are modeled thanks to a two-dimensional Hawkes
process in the following way: For $t\geq0$,
\[
P_t=N^+_t-N^-_t,
\]
where $(N^+,N^-)$ is a two-dimensional Hawkes process with intensity
\[
\pmatrix{ \lambda^+_t \vspace*{2pt}
\cr
\lambda^-_t } %
= %
\pmatrix{ \mu \vspace*{2pt}
\cr
\mu } %
+ \int_0^t %
\pmatrix{ \phi_1(t-s) & \phi_2(t-s) \vspace*{2pt}
\cr
\phi_2(t-s) & \phi_1(t-s) } %
\pmatrix{
dN^+_s \vspace*{2pt}
\cr
dN^-_s } %
,
\]
with $\phi_1$ and $\phi_2$, two nonnegative measurable functions such
that the stability condition
\[
\int_0^{+\infty} \phi_1(s)\,ds +\int
_0^{+\infty} \phi_2(s)\,ds<1
\]
is satisfied.

This model takes into account the discreteness and the
negative autocorrelation of prices at the microstructure level.
Moreover, it is shown in \cite{bacry2012scaling} that when one
considers this price at large time scales, the stability condition
implies that after suitable renormalization, it converges toward a
Brownian motion (with a given volatility).

\subsection{Scaling limit}

In the same spirit as in Section~\ref{s2}, we consider the scaling
limit of the Hawkes-based price process when the stability condition
becomes almost violated. More precisely, following the construction of
multivariate Hawkes processes of \cite{bacry2012scaling}, for every
observation interval $[0,T]$, we define the Hawkes process
$(N^{T+},N^{T-})$ with intensity
\[
\pmatrix{ \lambda^{T+}_t \vspace*{2pt}
\cr
\lambda^{T-}_t } %
= %
\pmatrix{ \mu
\vspace*{2pt}
\cr
\mu } %
+ \int_0^t
\pmatrix{ \phi_1^T(t-s) &
\phi_2^T(t-s) \vspace*{2pt}
\cr
\phi_2^T(t-s)
& \phi _1^T(t-s) } %
\pmatrix{dN^{T+}_s
\vspace*{2pt}
\cr
dN^{T-}_s } %
,
\]
with $\phi_1^T$ and $\phi_2^T$, two nonnegative measurable functions.
Note that in this construction, $N^{T+}$ and $N^{T-}$ do not have
common jumps; see \cite{bacry2012scaling} for details. We consider the
following assumption.

\begin{assumption}\label{h1p}
For $i=1,2$ and $t\in\mathbb{R}^+$,
\[
\phi^T_i(t)=a_T\phi_i(t),
\]
where $(a_T)_{T\geq0}$ is a sequence of positive numbers converging to
one such that for all $T$, $a_T<1$ and $\phi_1$ and $\phi_2$ are two
non negative measurable functions such that
\[
\int_0^{+\infty}\phi_1(s)+
\phi_2(s) \,ds=1\quad\mbox{and}\quad \int_0^{+\infty
}s
\bigl(\phi_1(s)+\phi_2(s) \bigr) \,ds=m<\infty.
\]
Moreover, the support of $\phi_2$ has non zero Lebesgue measure and
for $i=1,2$, $\phi_i$ is differentiable with derivative $\phi
_i'$ such that $\|\phi_i'\|_{\infty}<+\infty$ and $\|\phi_i'\|
_{1}<+\infty$.
\end{assumption}

We will also make the following technical assumption.

\begin{assumption}
\label{h2p}
Let
\[
\psi^T_+=\sum_{k\geq1}
\bigl(a_T(\phi_1+\phi_2)\bigr)^{*k}
\quad\mbox{and}\quad\rho ^T(x)=T\frac{\psi^T_+}{\|\psi^T_+\|_1}(Tx).
\]
There exists $K_\rho>0$ such that for all $x\geq0$ and $T>0$,
\[
\bigl|\rho^T(x)\bigr|\leq K_\rho.
\]
\end{assumption}

We work with the renormalized price process
%
\begin{equation}
\label{defP} P_t^T=\frac{1}{T}
\bigl(N^{T+}_{Tt}-N^{T-}_{Tt}\bigr).
\end{equation}

The following theorem states that if we consider the rescaled
price process over the right time interval, that is, if we take $T$ of
order $1/(1-\|\phi_1\|_1-\|\phi_2\|_1)$, it asymptotically behaves like
a Heston model; see \cite{heston1993closed}.

\begin{theorem}
\label{Th3}
Let $\phi=\phi_1-\phi_2$. Assume that \eqref{regime} holds. Under
Assumptions~\ref{h1p} and \ref{h2p},
the sequence of Hawkes-based price models $(P^T_t)$ converges in law,
for the Skorohod topology, toward a Heston-type process $P$ on $[0,1]$
defined by
\[
\cases{ %
\displaystyle dC_t=\biggl(
\frac{2\mu}{\lambda}-C_t\biggr)\frac{\lambda}{m} \,dt +\frac{1}{m}
\sqrt {C_t} \,dB^1_t, & \quad $C_0=0,$
\vspace*{2pt}\cr
\displaystyle dP_t=\frac{1}{1-\|\phi\|_1} \sqrt{C_t} \,dB^2_t,
& \quad $P_0=0,$}
\]
with $(B^1,B^2)$ a bidimensional Brownian motion.
\end{theorem}

\section{Proofs}\label{proofs}

We gather in this section the proofs of Theorem~\ref{Th1}, Proposition~\ref{formlambdamart}, Theorems \ref{main} and \ref{Th3}.
In the following, $c$ denotes a constant that may vary from line to line.
\subsection{Proof of Theorem \texorpdfstring{\protect\ref{Th1}}{2.1}}

Let $v\in[0,1]$. From Lemma~4 in \cite{bacry2012scaling}, we get
\[
\E\bigl[N^T_{Tv}\bigr]=\mu Tv+\mu\int_0^{Tv}
\psi^T(Tv-s)s\,ds
\]
and
\[
N^T_{Tv}-\E\bigl[N^T_{Tv}
\bigr]=M^T_{Tv}+\int_0^{Tv}
\psi^T(Tv-s)M^T_s\,ds.
\]
Thus, using that
\[
\bigl\|\psi^T\bigr\|_1=\frac{\|\phi^T\|_1}{1-\|\phi^T\|_1},
\]
we deduce
\begin{eqnarray*}
\frac{1-\|\phi^T\|_1}{T}\bigl(N^T_{Tv}-\E\bigl[N^T_{Tv}
\bigr]\bigr)&\leq&\frac{1-\|\phi^T\|
_1}{T}\bigl(1+\bigl\|\psi^T\bigr\|_1
\bigr)\mathop{\operatorname{sup}}_{t\in[0,T]}\bigl|M^T_t\bigr|
\\
&\leq&\frac
{1}{T}\mathop{\operatorname{sup}}_{t\in[0,T]}\bigl|M^T_t\bigr|.
\end{eqnarray*}
Now recall that $M^T$ is a square integrable martingale with quadratic
variation process $N^T$. Thus we can apply
Doob's inequality which gives
\[
\E\Bigl[\Bigl(\mathop{\operatorname{sup}}_{t\in[0,T]}M^T_t
\Bigr)^2\Bigr]\leq4\mathop{\operatorname{sup}}_{t\in
[0,T]}\E\bigl[
\bigl(M^T_t\bigr)^2\bigr]\leq4 \E
\bigl[N^T_T\bigr]\leq4 \mu\frac{T}{1-\|
\phi^T\|_1}.
\]
Therefore, we finally obtain
\[
\E \biggl[\mathop{\operatorname{sup}}_{v\in[0,1]} \biggl(\frac{1-\|\phi^T\|
_1}{T}
\bigl(N^T_{Tv}-\E\bigl[N^T_{Tv}\bigr]
\bigr) \biggr)^2 \biggr]\leq\frac{4\mu}{T(1-\|\phi^T\|_1)},
\]
which gives the result since $T(1-\|\phi^T\|_1)$ tends to infinity.

\subsection{Proof of Proposition \texorpdfstring{\protect\ref{formlambdamart}}{2.1}}
From the definition of $\lambda^T$, using the fact that $\phi$ is
bounded on $[0,t]$, we can write
\[
\lambda_t^T=\mu+\int_0^t
\phi^T(t-s) \,dM^T_s+\int_0^t
\phi^T(t-s) \lambda _s^T \,ds.
\]
We now recall the following classical lemma; see, for example, \cite
{bacry2012scaling} for a proof.

\begin{lemma}
If $f(t)=h(t)+\int_0^t \phi^T(t-s) f(s) \,ds$ with $h$ a measurable
locally bounded function, then
\[
f(t)=h(t)+\int_0^t \psi^T(t-s)
h(s) \,ds.
\]
\end{lemma}

We apply this lemma to the function $h$ defined by
\[
h(t)=\mu+\int_0^t \phi^T(t-s)
\,dM^T_s.
\]
Thus, we obtain
%
\begin{eqnarray}
\label{e1} \lambda_t^T&=&\mu+\int_0^t
\phi^T(t-s) \,dM^T_s
\nonumber
\\[-8pt]
\\[-8pt]
\nonumber
&&{}+\int_0^t
\psi^T(t-s) \biggl(\mu +\int_0^s
\phi^T(s-r) \,dM^T_r \biggr)\,ds.
\end{eqnarray}
Now remark that using Fubini theorem and the fact that
\[
\psi^T \ast\phi^T=\psi^T-\phi^T,
\]
we get
\begin{eqnarray*}
\int_0^t \psi^T(t-s) \int
_0^s \phi^T(s-r)
\,dM^T_r \,ds&=& \int_0^t
\int_0^t \mathrm{1}_{r\leq s}
\psi^T(t-s) \phi^T(s-r) \,ds \,dM^T_r
\\
&=& \int_{0}^t \int_0^{t-r}
\psi^T(t-r-s) \phi^T(s)\,ds \,dM^T_r
\\
&=& \int_{0}^t \psi^T \ast
\phi^T(t-r) \,dM^T_r
\\
&=& \int_{0}^t \psi^T (t-r)
\,dM_r^T-\int_{0}^t
\phi^T (t-r) \,dM_r^T.
\end{eqnarray*}
We conclude the proof rewriting \eqref{e1} using this last equality.

\subsection{Proof of Theorem \texorpdfstring{\protect\ref{main}}{2.2}}

Before starting the proof of Theorem~\ref{main}, we give some
preliminary lemmas.

\subsubsection{Preliminary lemmas}
We start with some lemmas on $\phi$ and its Fourier transform $\hat{\phi
}$ (the associated characteristic function).

\begin{lemma}
\label{H4}
Let $\delta>0$. There exists $\varepsilon>0$ such that for any real
number $z$ with $|z|\geq\delta$,
\[
\bigl|1-\hat{\phi}(z)\bigr|\geq\varepsilon.
\]
\end{lemma}

\begin{pf}
Since $\phi$ is bounded, $\hat{\phi}(z)$ tends to zero as $z$ tends to
infinity. Consequently,
there exists $b>\delta$ such that for all $z$ such that $|z|\geq b$,
\[
\bigl|\hat{\phi}(z)\bigr|\leq\tfrac{1}{2}.
\]
Now, let $M$ denote the supremum of the real part of $\hat{\phi}$ on
$[-b,-\delta]\cup[\delta,b]$, since $\hat\phi$ is continuous this
supremum is attained at some point $z_0$. We have $M=\operatorname{Re}(\hat{\phi
}(z_0))=\E[\cos(z_0X)]$, with $X$ a random variable with density $\phi
$. Since $\phi$ is continuous, almost surely, $X$ does not belong to
$2\pi/z_0 \mathbb{Z}$. Thus $M=\E[\cos(z_0X)]<1$.
Therefore, taking $\varepsilon=\min(\frac{1}{2},1-M)$ we have the lemma.
\end{pf}

Using that $\|\phi'\|_1<+\infty$, integrating by parts, we
immediately get the following lemma.

\begin{lemma}
\label{l43}
Let $z\in\mathbb{R}$. We have $|\hat{\phi}(z)|\leq c/|z|$.
\end{lemma}

We now turn to the function $\rho^T$ defined in \eqref{rho}.
We have the following result.

\begin{lemma}
\label{brho}
There exist $c>0$ such that for all real $z$ and $T\geq1$,
\[
\bigl|\widehat{\rho}^T(z)\bigr|\leq c\biggl(1\wedge\biggl|\frac{1}{z}\biggr|
\biggr).
\]
\end{lemma}

\begin{pf}
First note that as the Fourier transform of a random variable,
\mbox{$|\widehat{\rho}^T|\leq1$}.
Furthermore, using Lemma~\ref{H4} together with the fact that
\[
\int_0^{+\infty} x\phi(x) \,dx=m<+\infty,
\]
we get that there exist $\delta>0$ and $\varepsilon>0$ such that if
$|x|\leq\delta$,
\[
\bigl|\operatorname{Im}(\hat{\phi})(x)\bigr|\geq\frac{m}{2}|x|
\]
and if $|x|\geq\delta$,
\[
\bigl|1-\hat{\phi}(x)\bigr|\geq\varepsilon.
\]
Therefore, we deduce
that if $|z/T|\leq\delta$,
\[
\bigl|\widehat{\rho}^T(z)\bigr|=\biggl|\frac{(1-a_T)
\hat{\phi}({z}/{T})}{1-a_T \hat
{\phi}({z}/{T})}\biggr|\leq\frac{(1-a_T)}{a_T|\operatorname{Im}(\hat{\phi})
(
{z}/{T})|}
\leq\frac{2(1-a_T)T}{a_Tm|z|}\leq c/|z|
\]
and, thanks to Lemma~\ref{l43}, if $|z/T|\geq\delta$
\[
\bigl|\widehat{\rho}^T(z)\bigr|\leq\frac{(1-a_T)|\hat{\phi}({z}/{T})|}{|1-
\hat{\phi}({z}/{T})|}\leq\frac
{c(1-a_T)T}{|z|\varepsilon}
\leq c/|z|.
\]
\upqed\end{pf}

The next lemma gives us the $L^2$ convergence of $\rho^T$.

\begin{lemma}
\label{pi}
Let $\rho(x)=\frac{\lambda}{m}e^{-{x\lambda}/{m}}$ be the density
of the exponential random variable with parameter $\lambda/m$. We have
the following convergence, where $|\cdot|_2$ denotes the $L^2$ norm on
$\mathbb{R}^+$:
\[
\bigl|\rho^T-\rho\bigr|_2\rightarrow0.
\]
\end{lemma}

\begin{pf}
Using the Fourier isometry, we get
\[
\bigl|\rho^T-\rho\bigr|_2=\frac{1}{2\pi}\bigl|\widehat{
\rho}^T-\widehat{\rho}\bigr|_2.
\]
From Proposition~\ref{renyi}, for given $z$, we have $(\widehat{\rho}
^T(z)-\widehat{\rho}(z))\rightarrow0$.
Thanks to Lemma~\ref{brho}, we can apply the dominated convergence
theorem which gives that this convergence also takes place in $L^2$.
\end{pf}

We now give a Lipschitz type property for $\rho^T$.

\begin{lemma}
\label{lrho}
There exists $c>0$ such that for all $x\geq0$, $y\geq0$ and $T \geq1$,
\[
\bigl|\rho^T(x)-\rho^T(y)\bigr|\leq cT |x-y|.
\]
\end{lemma}

\begin{pf}
We simply compute the derivative of $\rho^T$ on $\mathbb{R}_+$, which
is given by
\[
\bigl(\rho^T\bigr)'(x)=T\biggl(\phi'(Tx)
\frac{T}{\|\psi^T\|_1}+\phi'\ast\rho ^T(Tx)\biggr).
\]
Using that $\|\psi^T\|_1=a_T/(1-a_T)$ together with the fact that
$T(1-a_T)\rightarrow\lambda$, we get
\[
\bigl|\bigl(\rho^T\bigr)'(x)\bigr|\leq T \bigl(c\bigl\|
\phi'\bigr\|_\infty+\bigl\|\phi'\bigr\|_1\bigl\|
\rho^T\bigr\|_\infty\bigr).
\]
\upqed\end{pf}

We now consider the function $f^T$ defined for $x\geq0$ by
\[
f^T(x)=\frac{m}{\lambda}\frac{a_T}{u_T}\rho^T(x)-e^{-{x}/{d_0}}.
\]
We have the following obvious corollaries.

\begin{corollary}
\label{pif}
We have
\[
\int\bigl|f^T(x)\bigr|^2\,dx\rightarrow0.
\]
\end{corollary}

\begin{corollary}
\label{fbound}
There exists $c>0$ such that for any $z\geq0$,
\[
\bigl|f^T(z)\bigr|\leq c.
\]
\end{corollary}

\begin{corollary}
\label{cvd}
There exists $c>0$ such that for any $z\geq0$,
\[
\bigl|\widehat{f}^T(z)\bigr|\leq c\biggl(\biggl|\frac{1}{z}\biggr|\wedge1\biggr).
\]
\end{corollary}

\begin{corollary}
\label{flip}
There exists $c>0$ such that for all $x\geq0$, $y\geq0$ and $T\geq1$,
\[
\bigl|f^T(x)-f^T(y)\bigr|\leq cT |x-y|.
\]
\end{corollary}

We finally give a lemma on the integrated difference
associated to the function~$f^T$.

\begin{lemma}
\label{regeps}
For any $0<\varepsilon<1$, there exists $c_\varepsilon$ such that for
all $t, s\geq0$,
\[
\int_\mathbb{R} \bigl(f^T(t-u)-f^T(s-u)
\bigr)^2\,du\leq c_\varepsilon |t-s|^{1-\varepsilon}.
\]
\end{lemma}

\begin{pf}
Defining $g_{t,s}^T(u)=f^T(t-u)-f^T(s-u)$, we easily get
\[
\bigl|\widehat{g}_{t,s}^T(w)\bigr|=\bigl|e^{-iwt}-e^{-iws}\bigr|
\bigl|\widehat{f}^T(w)\bigr|.
\]
Thus, from Corollary~\ref{cvd} together with the fact that
\[
\biggl|\frac{e^{-iwt}-e^{-iws}}{w(t-s)}\biggr|\leq1,
\]
we get
\begin{eqnarray*}
&&\int_\mathbb{R} \bigl(f^T(t-u)-f^T(s-u)
\bigr)^2\,du\\
&&\qquad\leq c\int_\mathbb{R} \bigl|\widehat{g}_{t,s}^T(w)\bigr|^2\,dw
\\
&&\qquad\leq c\int_\mathbb{R} \bigl|e^{-iwt}-e^{-iws}\bigr|^{2}
\biggl(\biggl|\frac{1}{w^2}\biggr|\wedge 1\biggr)\,dw
\\
&&\qquad\leq c\int_\mathbb{R} 2^{1+\varepsilon}\biggl|\frac
{e^{-iwt}-e^{-iws}}{w(t-s)}\biggr|^{1-\varepsilon}
\biggl(\biggl|\frac{1}{w^2}\biggr|\wedge 1\biggr)w^{1-\varepsilon}\,dw|t-s|^{1-\varepsilon}
\\
&&\qquad\leq c_\varepsilon|t-s|^{1-\varepsilon}.
\end{eqnarray*}
\upqed\end{pf}

\subsubsection{Proof of the first part of Theorem \texorpdfstring{\protect\ref{main}}{2.2}}

We now begin with the proof of the first assertion in Theorem~\ref
{main}. We split this proof into several steps.\vspace*{2pt}

\paragraph*{Step 1: Convenient rewriting of $C^T$}
In this step, our goal is to obtain a suitable expression for $C_t^T$.
Let $d_0=m/\lambda$. Inspired by the limiting behavior of $\psi^T$
given in Proposition~\ref{renyi}, we write equation \eqref{C2} under
the form
\[
C_t^T=R^T_t+\mu
\bigl(1-e^{-{t}/{d_0}}\bigr)+\frac{\sqrt{\lambda}}{m}\int_0^t
e^{-{(t-s)}/{d_0}} \sqrt{C_s^T}
\,dB^T_s,
\]
where $R^T_t$ is obviously defined. Using integration by parts (for
finite variation processes), we get
\begin{eqnarray*}
C_t^T&=&R^T_t+\frac{\mu}{d_0}
\int_0^t e^{-{v}/{d_0}}\,dv+\frac{\sqrt {\lambda}}{m}
\int_0^t\sqrt{C_v^T}
\,dB^T_v\\
&&{}-\frac{\sqrt{\lambda}}{md_0}\int_0^t
\biggl(\int_0^v e^{-{(v-s)}/{d_0}} \sqrt
{C_s^T} \,dB^T_s
\biggr)\,dv.
\end{eqnarray*}
Then remarking that
\[
\frac{\sqrt{\lambda}}{md_0}\int_0^v e^{-{(v-s)}/{d_0}}
\sqrt{C_s^T} \,dB^T_s=
\frac{1}{d_0} \bigl(C_v^T-R^T_v-
\mu\bigl(1-e^{-{v}/{d_0}}\bigr) \bigr),
\]
we finally derive
%
\begin{equation}
\label{form3} C_t^T=U^T_t+
\frac{1}{d_0} \int_0^t \bigl(
\mu-C^T_s\bigr)\,ds +\frac{\sqrt{\lambda
}}{m}\int
_0^t \sqrt{C^T_s}
\,dB^T_s,
\end{equation}
with
\[
U^T_t=R^T_t+\frac{1}{d_0}
\int_0^tR^T_s\,ds.
\]
Form \eqref{form3} will be quite convenient in order to study the
asymptotic behavior of $C_t^T$. Indeed, we will show that $U^T_t$
vanishes so that $\eqref{form3}$ almost represents a stochastic
differential equation.\vspace*{2pt}

\paragraph*{Step 2: Preliminaries for the convergence of $U^T$}

We now want to prove that the sequence of processes $(U^T_t)_{t\in
[0,1]}$ converges to zero in law, for the Skorohod topology, and
therefore uniformly on compact sets on $[0,1]$ (u.c.p.). We show here that
to do so, it is enough to study a (slightly) simpler process than
$U^T$. First, it is clear that showing the convergence of
$(R^T_t)_{t\in[0,1]}$ to zero gives also the convergence of $U^T$. Now
recall that
\begin{eqnarray*}
R^T_t&=&\mu(1-a_T)-\mu\biggl(
\bigl(1-e^{-{t}/{d_0}}\bigr)- \int_0^{t}a_T
T\frac{\psi
^T}{\|\psi^T\|_1}(Ts)\,ds\biggr)\\
&&{}+\sqrt{\lambda}\int_0^t
\biggl(\psi ^T\bigl(T(t-s)\bigr)-\frac{1}{m}e^{-{(t-s)}/{d_0}}
\biggr) \sqrt{C_t^T} \,dB^T_s.
\end{eqnarray*}
Since $a_T$ tends to one, the first term tends to zero. For $t\in
[0,1]$, Proposition~\ref{renyi} gives us the convergence of
\[
\int_0^{t}a_T T
\frac{\psi^T}{\|\psi^T\|_1}(Ts)\,ds
\]
toward $1-e^{-{t}/{d_0}}$. Using Dini's theorem, we get that this
convergence is in fact uniform over $[0,1]$. Thus, using equation \eqref
{defB}, we see that it remains to show that $(Y_t^T)_{t\in[0,1]}$ goes
to zero, with
\[
Y_t^T=\int_0^t
\bigl(m\psi^T\bigl(T(t-u)\bigr)-e^{-{(t-u)}/{d_0}}\bigr)\,d
\overline{M}^T_t,
\]
where $\overline{M}^T_t=M^T_{tT}/T$.\vspace*{2pt}

\paragraph*{Step 3: Finite dimensional convergence of $Y^T$}

We now show the finite dimensional convergence of $(Y^T_t)_{t\in[0,1]}$.

\begin{lemma}
\label{fidi}
For any $(t_1,\ldots,t_n)\in[0,1]^n$, we have the following convergence in law:
\[
\bigl(Y^T_{t_1},\ldots,Y^T_{t_n}
\bigr)\rightarrow0.
\]
\end{lemma}

\begin{pf}
First note that the quadratic variation of $\overline{M}^T$ at time $t$
is given by $N^T_{tT}/T^2$, whose predicable compensator process at
time $t$ is simply equal to
\[
\frac{1}{T^2}\int_0^{tT}
\lambda_s^T\,ds.
\]
Using this together with the fact that
\[
\E\bigl[\lambda_t^T\bigr]=\mu+\mu\int
_0^t\psi^T(t-s)\,ds\leq\mu+\mu
\frac
{a_T}{1-a_T}\leq cT,
\]
we get
\[
\E\bigl[\bigl(Y_t^T\bigr)^2\bigr]\leq c\int
_0^t \bigl(m\psi^T\bigl(T(t-s)
\bigr)-e^{-{(t-s)}/{d_0}}\bigr)^2\,ds.
\]
Now remark that
\[
m\psi^T\bigl(T(t-s)\bigr)-e^{-{(t-s)}/{d_0}}=f^T(t-s),
\]
where $f^T$ is defined by $f^T(x)=0$ for $x< 0$ and
\[
f^T(x)=\frac{m}{\lambda}\frac{a_T}{u_T}\rho^T(x)-e^{-{x}/{d_0}}
\]
for $x\geq0$, with $\rho^T$ the function introduced in equation \eqref{rho}.
From Corollary~\ref{pif},
\[
\E\bigl[\bigl(Y_t^T\bigr)^2\bigr]
\rightarrow0,
\]
which gives the result.\vspace*{2pt}
\end{pf}

\paragraph*{Step 4: A Kolmogorov-type inequality for $Y^T$}

To prove the convergence of $Y^T$ toward $0$, it remains to show its
tightness. We have the following Kolmogorov-type inequality on the
moments of the increments of $Y^T$, which is a first step in order to
get the tightness.

\begin{lemma}
\label{kolm4}
For any $\varepsilon>0$, there exists $c_\varepsilon>0$ such that for
all $T\geq1$, $0\leq t,s\leq1$,
%
\begin{equation}
\label{kolm4res} \E\bigl[\bigl(Y_t^T-Y_s^T
\bigr)^4\bigr]\leq c_\varepsilon\biggl(|t-s|^{3/2-\varepsilon}+
\frac
{1}{T^2}|t-s|^{1-\varepsilon}\biggr).
\end{equation}
\end{lemma}

\begin{pf}
Let $\mu^{\E[M^T_4]}$ denote the fourth moment measure of $M^T$; see
the Appendix in \cite{appendix} for definition and properties. We have
\begin{eqnarray*}
&&\E\bigl[\bigl(Y_t^T-Y_s^T
\bigr)^4\bigr]\\
&&\qquad=\frac{1}{T^4}\int_{[0,T]^4}
\Biggl(\prod_{i=1}^4 \biggl[f^T
\biggl(t-\frac{t_i}{T}\biggr)-f^T\biggl(s-\frac{t_i}{T}
\biggr) \biggr] \Biggr) \mu^{\E
[M^T_4]}(dt_1,dt_2,dt_3,dt_4).
\end{eqnarray*}
Therefore, using Lemma A.17 in \cite{appendix}, we obtain
\begin{eqnarray*}
\E\bigl[\bigl(Y_t^T-Y_s^T
\bigr)^4\bigr] &\leq&\frac{c}{T^3}\int_0^T
\biggl|f^T\biggl(t-\frac
{u}{T}\biggr)-f^T\biggl(s-
\frac{u}{T}\biggr)\biggr|^4 \,du
\\
&&{}+ \frac{c}{T^3}\int_0^T
\biggl|f^T\biggl(t-\frac{u}{T}\biggr)-f^T\biggl(s-
\frac{u}{T}\biggr)\biggr|^3 \,du\\
&&\quad{}\times \int_0^T
\biggl|f^T\biggl(t-\frac{u}{T}\biggr)-f^T\biggl(s-
\frac{u}{T}\biggr)\biggr|\,du
\\
&&{}+ \frac{c}{T^2}\int_0^T
\biggl|f^T\biggl(t-\frac{u}{T}\biggr)-f^T\biggl(s-
\frac{u}{T}\biggr)\biggr|^2 \,du\\
&&\quad{}\times\int_0^T
\biggl|f^T\biggl(t-\frac{u}{T}\biggr)-f^T\biggl(s-
\frac{u}{T}\biggr)\biggr|^2\,du
\\
&&{}+ \frac{c}{T^3} \biggl(\int_0^T
\biggl|f^T\biggl(t-\frac{u}{T}\biggr)-f^T\biggl(s-
\frac{u}{T}\biggr)\biggr| \,du \biggr)^2\\
&&\quad{}\times\int_0^T
\biggl|f^T\biggl(t-\frac{u}{T}\biggr)-f^T\biggl(s-
\frac{u}{T}\biggr)\biggr|^2\,du.
\end{eqnarray*}
Then, using the Cauchy--Schwarz inequality together with Corollary~\ref
{fbound} and Lemma~\ref{regeps}, we get
\[
\int_0^T \biggl|f^T\biggl(t-
\frac{u}{T}\biggr)-f^T\biggl(s-\frac{u}{T}\biggr)\biggr| \,du
\leq c_{\varepsilon
}T \sqrt{|t-s|^{1-\varepsilon}}
\]
and for $p=2,3,4$,
\[
\int_0^T \biggl|f^T\biggl(t-
\frac{u}{T}\biggr)-f^T\biggl(s-\frac{u}{T}
\biggr)\biggr|^p \,du\leq c_{\varepsilon
}T |t-s|^{1-\varepsilon},
\]
which allows us to complete the proof.\vspace*{2pt}
\end{pf}

\paragraph*{Step 5: Tightness}

Let us define $\tilde{Y}^T$ the linear interpolation of $Y^T$ with mesh $1/T^4$,
\[
\tilde{Y}^T_t=Y^T_{{ \lfloor tT^4  \rfloor
}/{T^4}}+
\bigl(tT^4- \bigl\lfloor tT^4 \bigr\rfloor\bigr)
\bigl(Y^T_{{ (\lfloor
tT^4  \rfloor+1)}/{T^4}}-Y^T_{{ \lfloor tT^4  \rfloor}/{T^4}}\bigr).
\]
We use this interpolation since for $t-s=1/T^4$, both terms on the
right-hand side of \eqref{kolm4res} have the same order of magnitude
and for $t-s>1/T^4$ the second term becomes negligible. We have the
following lemma.

\begin{lemma}\label{tight}
The sequence $(\tilde{Y}^T)$ is tight.
\end{lemma}

\begin{pf}
We want to apply the classical Kolmogorov tightness criterion (see~\cite
{billingsley2009convergence}) that states that if there exist
$\gamma> 1$ and $c > 0$ such that for any $0\leq s\leq t\leq1$,
\[
\E\bigl|\tilde{Y}^T_t-\tilde{Y}^T_s\bigr|^4
\leq c|t-s|^\gamma,
\]
then $\tilde{Y}^T$ is tight.
Note that such inequality can of course not hold for $Y^T$ since it is
not continuous.
Let $n_t^T= \lfloor tT^4  \rfloor$ and $n_s^T= \lfloor
sT^4  \rfloor$. Let $0<\varepsilon,\varepsilon'\leq1/4$ and
$T\geq1$. There are three cases:
\begin{itemize}
\item If $n_t^T=n_s^T$, using Lemma~\ref{kolm4}, we
obtain that
\[
\E\bigl[\bigl(\tilde{Y}^T_t-\tilde{Y}^T_s
\bigr)^4\bigr]
\]
is smaller than
\begin{eqnarray*}
|t-s|^4 T^{16} \E\bigl[(Y_{{(n_t^T+1)}/{T^4}}-Y_{{n_t^T}/{T^4}})^4
\bigr] &\leq& c_\varepsilon\frac{1}{T^{4(3/2-\varepsilon)}} T^{16}
|t-s|^{4}\\
&\leq& c_\varepsilon\frac{1}{T^{4(3/2-\varepsilon)}} T^{16}
|t-s|^{1+\varepsilon'}\frac{1}{T^{4(3-\varepsilon')}}.
\end{eqnarray*}
Since $0<\varepsilon,\varepsilon'\leq1/4$, this leads to
\[
\E\bigl[\bigl(\tilde{Y}^T_t-\tilde{Y}^T_s
\bigr)^4\bigr]\leq c_\varepsilon |t-s|^{1+\varepsilon'}.
\]

\item If $n_t^T=n_s^T+1$,
\[
\E\bigl[\bigl(\tilde{Y}^T_t-\tilde{Y}^T_s
\bigr)^4\bigr] \leq c\E\bigl[\bigl(\tilde{Y}^T_t-
\tilde {Y}^T_{{n_t^T}/
{T^4}}\bigr)^4\bigr]+c\E\bigl[\bigl(
\tilde{Y}^T_{{n_t^T }/
{T^4}}-\tilde {Y}^T_s\bigr)^4
\bigr]\leq c_\varepsilon |t-s|^{1+\varepsilon'}.
\]

\item If $n_t^T\geq n_s^T+2$, using again Lemma~\ref
{kolm4}, we get
\begin{eqnarray*}
\E\bigl[\bigl(\tilde{Y}^T_t-\tilde{Y}^T_s
\bigr)^4\bigr]&\leq& c \E\bigl[\bigl(\tilde{Y}^T_t-
\tilde {Y}^T_{{n_t^T}/
{T^4}}\bigr)^4\bigr]+c\E\bigl[\bigl(
\tilde{Y}^T_{{(n_s^T+1)}/
{T^4}}-\tilde {Y}^T_s\bigr)^4
\bigr]\\
&&{}+c\E\bigl[\bigl(\tilde{Y}^T_{{n_t^T}/
{T^4}}-\tilde{Y}^T_{{(n_s^T+1)}/
{T^4}}\bigr)^4\bigr]
\\
&\leq &c_\varepsilon\biggl(\frac{1}{T^4}\biggr)^{1+\varepsilon'}+c_\varepsilon\biggl|
\frac
{n_t^T}{T^4}-\frac{n_s^T+1}{T^4}\biggr|^{{3}/{2}-\varepsilon}
\\
&\leq& c_\varepsilon|t-s|^{\operatorname{min}({3}/{2}-\varepsilon,1+\varepsilon')}.
\end{eqnarray*}

Hence the Kolmogorov criterion holds, which implies the
tightness of $\tilde{Y}^T$.\quad\qed
\end{itemize}\noqed
\end{pf}

We now show that the difference between $Y^T$ and $\tilde
{Y}^T$ tends uniformly to zero.

\begin{lemma}\label{local}
We have the following convergence in probability:
\[
\mathop{\sup}_{|t-s|\leq{1}/{T^4}}\bigl|Y_t^T-Y_s^T\bigr|
\rightarrow0.
\]
\end{lemma}

\begin{pf}
Recall that for $0\leq s\leq t\leq1$,
\[
\bigl|Y^T_t-Y_s^T\bigr|=\biggl|\int
_0^s f^T(t-u)-f^T(s-u)\,d
\overline{M}^T_u+\int_s^tf^T(t-u)\,d
\overline{M}^T_u\biggr|.
\]
Thus, we have that $|Y^T_t-Y_s^T|$ is smaller than
\begin{eqnarray*}
&&\int_0^{sT} \bigl|f^T(t-u/T)-f^T(s-u/T)\bigr|
\bigl(dN^T_u+\lambda_u \,du\bigr)
\frac
{1}{T}\\
&&\qquad+\int_{sT}^{tT}
\bigl|f^T(t-u/T)\bigr| \bigl(dN^T_u+
\lambda_u \,du\bigr)\frac{1}{T}.
\end{eqnarray*}
Using Corollaries \ref{fbound} and \ref{flip}, we obtain
\[
\bigl|Y^T_t-Y_s^T\bigr|\leq c|t-s|
\biggl(N_T^T+\int_0^T
\lambda_u^T\,du\biggr) + c\biggl(N^T_{tT}-N^T_{sT}+
\int_{sT}^{tT} \lambda^T_u
\,du\biggr)\frac{1}{T}.
\]
Consider now
\[
\mathop{\operatorname{sup}}_{|t-s|\leq1/T^4}\bigl|Y^T_t-Y_s^T\bigr|.
\]
This is smaller than
%
\begin{eqnarray}
\label{approx}&&c\frac{1}{T^4} \biggl(N_T^T+\int
_0^T \lambda _u^T\,du
\biggr)
\nonumber
\\[-8pt]
\\[-8pt]
\nonumber
&&\qquad{} + 2c\mathop{\operatorname {max}}_{i=0,\ldots, \lfloor T^4  \rfloor}\frac{1}{T}
\biggl(N^T_{({(i+1)}/{T^4})T}-N^T_{({i}/{T^4})T}+\int
_{({i}/{T^4})T}^{({(i+1)}/{T^4})T} \lambda^T_u \,du
\biggr).
\end{eqnarray}
From Lemma A.5 in \cite{appendix}, we have
\[
\E \biggl[N_T^T+\int_0^T
\lambda_u^T\,du \biggr]\leq cT^{2}.
\]
Thus, the first term on the right-hand side of \eqref{approx} tends to zero.
For the second term, we use Lemma A.15 in \cite{appendix} (with $t=\frac
{i+1}{T^4}T$ and $s=\frac{i}{T^4}T$) which gives that
\[
\E \biggl[ \biggl(\frac{1}{T}\biggl(N^T_{({(i+1)}/{T^4})T}-N^T_{({i}/{T^4})T}+
\int_{({i}/{T^4})T}^{({(i+1)}/{T^4})T} \lambda^T_u
\,du\biggr) \biggr)^3 \biggr]\leq \frac{c}{T^5}.
\]
So, for any $\varepsilon>0$, using Markov's inequality, we get
\[
\mathbb{P} \biggl[\frac{1}{T}\biggl(N^T_{({(i+1)}/{T^4})T}-N^T_
{({i}/{T^4})T}+
\int_{({i}/{T^4})T}^{({(i+1)}/{T^4)}T} \lambda^T_u
\,du\biggr)\geq \varepsilon \biggr] \leq\frac{c}{T^5 \varepsilon^3}.
\]
From this inequality, since the maximum is taken over a number of terms
of order~$T^4$, we easily deduce that the second term on the right-hand
side of \eqref{approx} tends to zero in probability.
\end{pf}

We end this step by the proposition stating the convergence
of $Y^T$.

\begin{proposition}
The process $Y^T$ converges u.c.p. to $0$ on $[0,1]$.
\end{proposition}

\begin{pf}
We have
\[
\sup_{t\in[0,1]}\bigl|Y^T_t\bigr|\leq\sup
_{t\in[0,1]}\bigl|\tilde{Y}^T_t\bigr|+\sup
_{t\in[0,1]}\bigl|\tilde{Y}_t^T-Y_t^T\bigr|.
\]
From Lemmas \ref{fidi} and \ref{tight} we get that $\tilde{Y}^T$
tends to zero, in law for the Skorohod topology. This implies the u.c.p.
convergence. Applying Lemma~\ref{local} we get the result.\vspace*{2pt}
\end{pf}

\paragraph*{Step 6: Limit of a sequence of SDEs}

In this last step, we show the convergence of the process $(C_t^T)_{t\in
[0,1]}$ toward a CIR process. To do so, we use the fact that
$C^T$ can almost be written under the form of a stochastic differential
equation. Indeed, recall that
\[
C_t^T=U^T_t+\frac{1}{d_0}
\int_0^t \bigl(\mu-C^T_s
\bigr)\,ds +\frac{\sqrt{\lambda
}}{m}\int_0^t \sqrt
{C^T_s} \,dB^T_s,
\]
with
\[
B_t^T=\frac{1}{\sqrt{T}}\sqrt{u_T}\int
_0^{tT} \frac{dM_s^T}{\sqrt {\lambda^T_s}}.
\]
Then we aim at applying Theorem~5.4 in \cite{kurtz1991weak} to $C^T$.
This result essentially says that for a sequence of SDEs where the
functions and processes defining the equations satisfy some convergence
properties, the laws of the solutions of the SDEs converge to the law
of the solution of the limiting SDE. We now check these convergence
properties.

The sequence of processes $(B^T)$ is a sequence of
martingales with jumps uniformly bounded by $c/\sqrt{\mu}$.
Furthermore, for $t\in[0,1]$,
the quadratic variation of $(B^T)$ at point $t$ is equal to
\[
\frac{u_T}{T}\int_0^{tT}
\frac{dN_s^T}{\lambda^T_s}=u_T \biggl(t+\int_0^{tT}
\frac{dM_s^T}{T\lambda^T_s} \biggr).
\]
Now, remark that
\[
\E \biggl[\biggl(\int_0^{tT}\frac{dM_s^T}{T\lambda^T_s}
\biggr)^2 \biggr]\leq\E \biggl[\int_0^{T}
\frac{1}{T^2\lambda^T_s}\,ds \biggr]\leq c/(T\mu).
\]
Therefore, we get that for any $t\in[0,1]$, the quadratic variation of
$(B^T)$ at point $t$ converges in probability to $t$. Thus, we can
apply Theorem VIII.3.11 in \cite{jacod1987limit} to deduce that
$(B^T_t)_{t\in[0,1]}$ converges in law for the Skorohod topology toward
a Brownian motion.

Since $U^T$ converges to a deterministic limit, we get the
convergence in law, for the product topology, of the couple
$(U^T_t,B^T_t)_{\in[0,1]}$ to $(0,B_t)_{\in[0,1]}$, with $B$ a Brownian
motion. The components of $(0,B_t)$ being continuous, the last
convergence also takes place for the Skorohod topology on the product
space.

Finally, recall that the (CIR) stochastic differential equation
\[
X_t=\int_0^t (
\mu-X_s)\frac{1}{d_0} \,ds +\frac{\sqrt{\lambda}}{m}\int
_0^t \sqrt{X_s} \,dB_s
\]
admits a unique strong solution on $[0,1]$. This, together with the
preceding elements enables us to readily apply Theorem~5.4 in \cite
{kurtz1991weak} to the sequence $C^T$, which gives the result.

\subsubsection{Proof of the second part of Theorem \texorpdfstring{\protect\ref{main}}{2.2}}

We now give the proof of the second part of Theorem~\ref{main} which
deals with the sequence of Hawkes processes~$N^T$.
Let
\[
V_t^T=\frac{(1-a_T)}{T}N^T_{tT}.
\]
We write
\[
V_t^T=\int_0^t
C_s^T\,ds+\hat{M}^T_t,
\]
where
\[
\hat{M}^T_t=\frac{(1-a_T)}{T}\biggl(N_{tT}^T-
\int_0^{tT} \lambda^T_s
\,ds\biggr)
\]
is a martingale.
Using Doob's inequality, we obtain
\[
\E\Bigl[\Bigl(\operatorname{sup}_{t\in[0,1]}\hat{M}^T_t
\Bigr)^2\Bigr]\leq4\E\bigl[\bigl(\hat{M}^T_1
\bigr)^2\bigr]\leq 4\biggl(\frac{(1-a_T)}{T}\biggr)^2 \E
\bigl[N_T^T\bigr] \leq\frac{4\mu(1-a_T)}{T} \rightarrow0.
\]
Moreover, $(C^T,t)$ converges in law over $[0,1]$ to $(C,t)$ for the
Skorokod topology.
This last remark and Theorem~2.6 in \cite{jakubowski1989convergence} on
the limit of sequences of stochastic integrals give the result.

\subsection{Proof of Theorem \texorpdfstring{\protect\ref{Th3}}{3.1}}

We first introduce some notation. In this proof, we write
\[
\phi^T=\phi_1^T-\phi_2^T\quad
\mbox{and}\quad\psi^T=\sum_{k=1}^{+\infty}
\bigl(\phi^T\bigr)^{*k}.
\]
Moreover, we set
\[
C^T_t=\frac{\lambda^{T+}_{tT}+\lambda^{T-}_{tT}}{T}
\]
and define
\[
\bigl(B^1\bigr)^T_t=\int
_0^{tT} \frac{dM^{T+}_s+dM^{T-}_s}{\sqrt{T(\lambda
^{T+}_{s}+\lambda^{T-}_{s})}},\qquad
\bigl(B^2\bigr)^T_t=\int_0^{tT}
\frac
{dM^{T+}_s-dM^{T-}_s}{\sqrt{T(\lambda^{T+}_{s}+\lambda^{T-}_{s})}},
\]
with
\[
M^{T+}_s=N^{T+}_s-\int
_0^s\lambda^{T+}_{s}\,ds, \qquad  M^{T-}_s=N^{T-}_s-
\int_0^s\lambda^{T-}_{s}\,ds.
\]
Finally, we set
\[
\overline{M}_t^{T+}=\frac{M_{Tt}^{T+}}{T},\qquad
\overline{M}_t^{T-}=\frac
{M_{Tt}^{T-}}{T}.
\]

We split the proof of Theorem~\ref{Th3} into several steps.

\subsubsection*{Step 1: Convenient rewriting}

In this first step, we rewrite the price, intensity and martingale
processes under more convenient forms. We have
\[
\lambda^{T+}_t-\lambda^{T-}_t=\int
_0^t \phi^T(t-s) \bigl(\lambda
^{T+}_s-\lambda^{T-}_s\bigr)\,ds+\int
_0^t \phi^T(t-s)
\bigl(dM^{T+}_s-dM^{T-}_s\bigr).
\]
Therefore, in the same way as in the proof of Proposition~\ref
{formlambdamart}, we get
\[
\lambda^{T+}_t-\lambda^{T-}_t=\int
_0^t\psi^T(t-s)
\bigl(dM^{T+}_s-dM^{T-}_s\bigr).
\]
From this last expression, we easily obtain
%
\begin{equation}
\label{eqdN} N^{T+}_t-N^{T-}_t=
\int_0^t \bigl(1+\Psi^T(t-u) \bigr)
\bigl(dM^{T+}_u-dM^{T-}_u\bigr),
\end{equation}
with
\[
\Psi^T(x)=\int_0^x
\psi^T(s) \,ds.
\]
Finally, note that
%
\begin{equation}
\label{eqhest} \overline{M}_t^{T+}-\overline{M}_t^{T-}=
\frac
{1}{T}\bigl(M_{Tt}^{T+}-M_{Tt}^{T-}
\bigr)=\int_0^t \sqrt{C^T_s}
\,d\bigl(B^2\bigr)^T_s.
\end{equation}

\subsubsection*{Step 2: Preliminary result}
For $s\in[0,1]$, we define
\[
X^T_s=\frac{\lambda_{sT}^{T+}-\lambda_{sT}^{T-}}{T}.
\]
We have the following important result.

\begin{lemma}\label{convX}
The process $X^T$ converges u.c.p. to $0$ on $[0,1]$.
\end{lemma}

\begin{pf}
We write
\[
X^T_t=\int_0^t
f^T_1(t-s)\,d\bigl(\overline{M}_s^{T+}-
\overline{M}_s^{T-}\bigr),
\]
with $f^T_1(x)=\psi^T(Tx)$. Note that Corollaries \ref{pif}, \ref
{fbound}, \ref{cvd} and \ref{flip} are valid
if in their statement, $f^T$ is replaced by $f^T_1$. In the proof of
Theorem~\ref{main}, we have shown the convergence to zero of the process
\[
Y^T_t=\int_0^T
f^T(t-s)\,d\overline{M}^T_s.
\]
Therefore, applying the same strategy but replacing $f^T$ by $f^T_1$
and $\overline{M}^T$ by $\overline{M}^{T+}-\overline{M}^{T-}$, it is
clear that we get the result.
\end{pf}

\subsubsection*{Step 3: Convergence of $(B^1,B^2)$}
In this step, we prove the convergence of $(B^1,B^2)$ toward a
two-dimensional Brownian motion. To do so, we study the quadratic
(co-)variations of the processes. Let $i\in\{1,2\}$, $j\in\{1,2\}$. We
denote by $[(B^i)^T,(B^j)^T]_t$ the quadratic co-variation of $B^i$ and
$B^j$ at time $t$.

\begin{lemma}
\label{B1B2}
We have the following convergence in probability:
\[
\bigl[\bigl(B^i\bigr)^T,\bigl(B^j
\bigr)^T\bigr]_t\rightarrow t\mathrm{1}_{i=j}.
\]
\end{lemma}

\begin{pf}
There are three cases:
\begin{itemize}
\item If $i=j=1$, using that $N^{T+}$ and $N^{T-}$ have
no common jumps, we get
\[
\bigl[\bigl(B^1\bigr)^T,\bigl(B^1
\bigr)^T\bigr]_t=\int_0^{tT}
\frac{dN_s^{T+}+dN_s^{T-}}{T(\lambda
^{T+}_s+\lambda^{T-}_s)}=t+\int_0^{tT}
\frac
{dM_s^{T+}+dM_s^{T-}}{T(\lambda^{T+}_s+\lambda^{T-}_s)}.
\]
Furthermore,
\[
\E \biggl[ \biggl(\int_0^{tT}\frac{dM_s^{T+}+dM_s^{T-}}{T(\lambda
^{T+}_s+\lambda^{T-}_s)}
\biggr)^2 \biggr]\leq\frac{ct}{T\mu}\rightarrow0.
\]
Therefore we have the result for $i=j=1$.

\item If $i=j=2$, the proof goes similarly.

\item If $i=1$ and $j=2$,
\begin{eqnarray*}
\bigl[\bigl(B^1\bigr)^T,\bigl(B^2
\bigr)^T\bigr]_t&=&\int_0^{tT}
\frac{dN_s^{T+}-dN_s^{T-}}{T(\lambda
^{T+}_s+\lambda^{T-}_s)}\\
&=&\int_0^{tT}\frac{dM_s^{T+}-dM_s^{T-}+\lambda
_s^{T+}\,ds-\lambda_s^{T-}\,ds}{T(\lambda^{T+}_s+\lambda^{T-}_s)}.
\end{eqnarray*}
As for the case $i=j=1$, we easily get
\[
\int_0^{tT}\frac{dM_s^{T+}-dM_s^{T-}}{T(\lambda^{T+}_s+\lambda
^{T-}_s)}\rightarrow0.
\]
It remains to show the convergence to zero of $Z_t^T$ defined by
\[
Z_t^T=\int_0^t
\frac{X^T_s}{C^T_s}\,ds.
\]
For any $\varepsilon>0$, we have
\[
\bigl|Z^T_t\bigr|\leq\int_0^t
\biggl(1\wedge\biggl|\frac{X^T_s}{\varepsilon}\biggr|\biggr) \,ds+\int_0^t
\mathrm{1}_{C^T_s< \varepsilon} \,ds.
\]
From Lemma~\ref{convX}, we have the convergence of the process $X^T$ to
zero. Furthermore, in
Lemma~\ref{cirbro} we will show that $C^T$ converge in law over $[0,1]$
toward a CIR process denoted
by $C$. Therefore, since the limiting processes are continuous, we have
the joint convergence
of $(X_T,C^T)$ to $(0,C)$. We now use Skorohod representation theorem
(without changing notation). Almost surely, for $T$ large enough, we have
\[
\mathop{\operatorname{sup}}_{s\in[0,1]}\bigl|X_s^T\bigr|\leq
\varepsilon^2,\qquad  \mathop{\operatorname{sup}}_{s\in[0,1]}\bigl|C_s^T-C_s\bigr|
\leq\varepsilon.
\]
This implies
\[
\int_0^t\biggl(1\wedge\biggl|\frac{X^T_s}{\varepsilon}\biggr|
\biggr) \,ds+\int_0^t\mathrm {1}_{C^T_s< \varepsilon}
\,ds\leq\varepsilon+\int_0^1\mathrm
{1}_{C_s<2\varepsilon} \,ds.
\]
Recall that the set of zeros of a CIR process on a finite time interval
has zero Lebesgue measure. Thus, using the dominated convergence
theorem, we easily see that choosing $\varepsilon$ conveniently, the
second term in the preceding inequality can be made arbitrarily small,
which completes the proof.\quad\qed
\end{itemize}
\noqed\end{pf}

Thus for any $T$, $(B^1)^T$ and $(B^2)^T$ are two martingales
with uniformly bounded jumps and their quadratic (co-)variations
satisfy Lemma~\ref{B1B2}. Consequently, Theorem VIII.3.11 of \cite
{jacod1987limit} gives us the following lemma.

\begin{lemma}
\label{MB2D}
We have
\[
\bigl(\bigl(B^1\bigr)^T,\bigl(B^2
\bigr)^T\bigr)\rightarrow\bigl(B^1,B^2\bigr),
\]
in law, for the Skorohod topology, where $(B^1,B^2)$ is a
two-dimensional Brownian motion.
\end{lemma}

\subsubsection*{Step 4: Convergence of $(C^T,(B^2)^T)$}
The aim of this step is to prove that the couple $(C^T,(B^2)^T)$
converges in law toward $(C,(B^2))$, with $C$ a CIR process and $B^2$ a
Brownian motion, independent of $C$. More precisely, we have the
following lemma.

\begin{lemma}
\label{cirbro} The couple of process $(C^T,(B^2)^T)$ converges in law,
for the Skorohod topology, over $[0,1]$, toward $(C,B^2)$, where $B^2$
is a Brownian motion independent of $C$ and $C$ is a CIR process satisfying
\[
C_t=\int_0^t \biggl(
\frac{2\mu}{\lambda}-C_s\biggr)\frac{\lambda}{m} \,ds +
\frac
{1}{m}\int_0^t
\sqrt{C_s} \,dW_s,
\]
with $W$ another Brownian motion, independent of $B^2$.
\end{lemma}

\begin{pf}
Let us consider the process $N^T=N^{T+}+N^{T-}$. It is a point process
with intensity
\[
\lambda^T_t=\lambda^{T+}_t+
\lambda^{T-}_t=2\mu+a_T\int_0^t
(\phi_1+\phi _2) (t-s)\,dN^T_s.
\]
Therefore, we are in the framework of Theorem~\ref{main}: $N^T$ is a
Hawkes process whose kernel has a norm that tends to $1$ at the right
speed and its renormalized intensity $C^T$ converges toward a CIR. Note
that the renormalizing factor here is $1/T$ and not $(1-a_T)$, which is
not an issue since \eqref{regime} holds. Thus we get the convergence of
$C^T$ toward a CIR.
To obtain the joint convergence, we just need to write the same proof
as for Theorem~\ref{main} (up to obvious changes), but using this time
Theorem~5.4 in \cite{kurtz1991weak} together with Lemma~\ref{MB2D}.
\end{pf}

\subsubsection*{Step 5: Technical results}
This fifth step consists in proving two technical results. The first
one is the following.

\begin{lemma}
\label{420}
The process
\[
R^T_t=\int_0^t \int
_{T(t-u)}^{+\infty}\psi^T(s)\,ds \,d\bigl(\overline
{M}_u^{T+}-\overline{M}_u^{T-}
\bigr)
\]
converges u.c.p. to $0$ on $[0,1]$.
\end{lemma}

\begin{pf}
We write
\[
R^T_t=\int_0^t
f^T_2(t-u) \,d\bigl(\overline{M}_u^{T+}-
\overline{M}_u^{T-}\bigr),
\]
with
\[
f^T_2(x)=\int_{Tx}^{+\infty}
\psi^T(s)\,ds.
\]
The result follows in the same way as in the proof of Lemma~\ref{convX}.
\end{pf}

We now give the last lemma of this step.

\begin{lemma}
\label{h5} We have
\[
\int_0^\infty\int_x^\infty
\phi_i(s)\,ds\,dx<\infty.
\]
\end{lemma}

\begin{pf}
Using integration by parts together with Assumption~\ref{h2p}, we get
\[
\int_0^\infty\int_x^\infty
\phi_i(s)\,ds\,dx=\int_0^\infty x \phi
_i(x)\,dx+\lim_{x\rightarrow\infty} x \int_x^\infty
\phi_i(s)\,ds\leq2m.
\]
\upqed\end{pf}

\subsubsection*{Step 6: End of the proof}
We finally show Theorem~\ref{Th3} in this step. Using \eqref{eqdN} we write
\begin{eqnarray*}
P^T_t& =&\biggl(1+\frac{\|\phi\|_1}{1-\|\phi\|_1}\biggr) \bigl(
\overline{M}_t^{T+}-\overline {M}_t^{T-}
\bigr)
\\
&&{}-\int_0^t \int_{T(t-u)}^{+\infty}
\psi^T(s)\,ds \,d\bigl(\overline {M}_u^{T+}-
\overline{M}_u^{T-}\bigr)\\
 &&{}-\biggl(\frac{\|\phi\|_1}{1-\|\phi\|_1}-
\frac{a_T\|\phi\|_1}{1-a^T\|\phi\|
_1}\biggr) \bigl(\overline{M}_t^{T+}-
\overline{M}_t^{T-}\bigr).
\end{eqnarray*}
Using Theorem~2.6 in \cite{jakubowski1989convergence} together with
Lemma~\ref{cirbro} and equation \eqref{eqhest}, we get the convergence
of the process
$\overline{M}^{T+}-\overline{M}^{T-}$, over $[0,1]$, for the Skorohod
topology, toward
\[
\int_0^t \sqrt{C_s}
\,dB^2_s.
\]
Moreover, in Lemma~\ref{420}, we have shown that the second term in the
decomposition of $P^T_t$ tends to zero.
Finally, the third term also vanishes since $\|\phi\|_1<1$. This
completes the proof.

\section*{Acknowledgment}
We thank Emmanuel Bacry for several interesting discussions.




\printaddresses


\begin{thebibliography}{41}


\bibitem{adamopoulos1976cluster}
\begin{barticle}[auto:STB|2014/02/12|14:17:21]
\bauthor{\bsnm{Adamopoulos},~\bfnm{L.}\binits{L.}}
(\byear{1976}).
\btitle{Cluster models for earthquakes: Regional comparisons}.
\bjournal{Journal of the International Association for Mathematical Geology}
\bvolume{8}
\bpages{463--475}.
\end{barticle}
\bptok{imsref}%
\endbibitem

\bibitem{ait2010modeling}
\begin{bmisc}[auto:STB|2014/02/12|14:17:21]
\bauthor{\bsnm{A{\"{\i}}t-Sahalia},~\bfnm{Y.}\binits{Y.}},
\bauthor{\bsnm{Cacho-Diaz},~\bfnm{J.}\binits{J.}} \AND
\bauthor{\bsnm{Laeven},~\bfnm{R.~J.}\binits{R.~J.}}
(\byear{2010}).
\bhowpublished{Modeling financial contagion using mutually exciting jump processes. Technical report,
National Bureau of Economic Research, Cambridge, MA}.
\end{bmisc}
\bptok{imsref}%
\endbibitem

\bibitem{alaya2012parameter}
\begin{barticle}[mr]
\bauthor{\bsnm{Alaya},~\bfnm{Mohamed~Ben}\binits{M.~B.}} \AND
\bauthor{\bsnm{Kebaier},~\bfnm{Ahmed}\binits{A.}}
(\byear{2012}).
\btitle{Parameter estimation for the square-root diffusions: Ergodic and nonergodic cases}.
\bjournal{Stoch. Models}
\bvolume{28}
\bpages{609--634}.
\bid{doi={10.1080/15326349.2012.726042}, issn={1532-6349}, mr={2995525}}
\end{barticle}
\bptok{imsref}%
\endbibitem

\bibitem{bacry2012non}
\begin{barticle}[auto:STB|2014/02/12|14:17:21]
\bauthor{\bsnm{Bacry},~\bfnm{E.}\binits{E.}},
\bauthor{\bsnm{Dayri},~\bfnm{K.}\binits{K.}} \AND
\bauthor{\bsnm{Muzy},~\bfnm{J.-F.}\binits{J.-F.}}
(\byear{2012}).
\btitle{Non-parametric kernel estimation for symmetric Hawkes processes. Application to high frequency financial data}.
\bjournal{Eur. Phys. J. B}
\bvolume{85}
\bpages{1--12}.
\end{barticle}
\bptok{imsref}%
\endbibitem

\bibitem{bacry2012scaling}
\begin{bmisc}[auto:STB|2014/02/12|14:17:21]
\bauthor{\bsnm{Bacry},~\bfnm{E.}\binits{E.}},
\bauthor{\bsnm{Delattre},~\bfnm{S.}\binits{S.}},
\bauthor{\bsnm{Hoffmann},~\bfnm{M.}\binits{M.}} \AND
\bauthor{\bsnm{Muzy},~\bfnm{J.~F.}\binits{J.~F.}}
(\byear{2012}).
\bhowpublished{Scaling limits for Hawkes processes and application to financial statistics.
Preprint. Available at \arxivurl{arXiv:1202.0842}}.
\end{bmisc}
\bptok{imsref}%
\endbibitem

\bibitem{bacry2013modelling}
\begin{barticle}[mr]
\bauthor{\bsnm{Bacry},~\bfnm{E.}\binits{E.}},
\bauthor{\bsnm{Delattre},~\bfnm{S.}\binits{S.}},
\bauthor{\bsnm{Hoffmann},~\bfnm{M.}\binits{M.}} \AND
\bauthor{\bsnm{Muzy},~\bfnm{J.~F.}\binits{J.~F.}}
(\byear{2013}).
\btitle{Modelling microstructure noise with mutually exciting point processes}.
\bjournal{Quant. Finance}
\bvolume{13}
\bpages{65--77}.
\bid{doi={10.1080/14697688.2011.647054}, issn={1469-7688}, mr={3005350}}
\end{barticle}
\bptok{imsref}%
\endbibitem

\bibitem{bacry2013hawkes}
\begin{bmisc}[auto:STB|2014/02/12|14:17:21]
\bauthor{\bsnm{Bacry},~\bfnm{E.}\binits{E.}} \AND
\bauthor{\bsnm{Muzy},~\bfnm{J.-F.}\binits{J.-F.}}
(\byear{2013}).
\bhowpublished{Hawkes model for price and trades high-frequency dynamics.
Preprint. Available at \arxivurl{arXiv:1301.1135}}.
\end{bmisc}
\bptok{imsref}%
\endbibitem

\bibitem{barczy2011asymptotic}
\begin{barticle}[mr]
\bauthor{\bsnm{Barczy},~\bfnm{M.}\binits{M.}},
\bauthor{\bsnm{Isp{\'a}ny},~\bfnm{M.}\binits{M.}} \AND
\bauthor{\bsnm{Pap},~\bfnm{G.}\binits{G.}}
(\byear{2011}).
\btitle{Asymptotic behavior of unstable {$\mathrm{INAR}(p)$} processes}.
\bjournal{Stochastic Process. Appl.}
\bvolume{121}
\bpages{583--608}.
\bid{doi={10.1016/j.spa.2010.11.005}, issn={0304-4149}, mr={2763097}}
\end{barticle}
\bptok{imsref}%
\endbibitem

\bibitem{bauwens2004dynamic}
\begin{bmisc}[auto:STB|2014/02/12|14:17:21]
\bauthor{\bsnm{Bauwens},~\bfnm{L.}\binits{L.}} \AND
\bauthor{\bsnm{Hautsch},~\bfnm{N.}\binits{N.}}
(\byear{2004}).
\bhowpublished{Dynamic latent factor models for intensity processes.
Working paper, UCL-CORE Center for Operations
Research and Econometrics}.
\end{bmisc}
\bptok{imsref}%
\endbibitem

\bibitem{billingsley2009convergence}
\begin{bbook}[mr]
\bauthor{\bsnm{Billingsley},~\bfnm{Patrick}\binits{P.}}
(\byear{1968}).
\btitle{Convergence of Probability Measures}.
\bpublisher{Wiley},
\blocation{New York}.
\bid{mr={0233396}}
\bptnote{check year}%
\end{bbook}
\bptok{imsref}%
\endbibitem

\bibitem{bouchaud2004fluctuations}
\begin{barticle}[auto:STB|2014/02/12|14:17:21]
\bauthor{\bsnm{Bouchaud},~\bfnm{J.-P.}\binits{J.-P.}},
\bauthor{\bsnm{Gefen},~\bfnm{Y.}\binits{Y.}},
\bauthor{\bsnm{Potters},~\bfnm{M.}\binits{M.}} \AND
\bauthor{\bsnm{Wyart},~\bfnm{M.}\binits{M.}}
(\byear{2004}).
\btitle{Fluctuations and response in financial markets: The subtle nature of ``random'' price changes}.
\bjournal{Quant. Finance}
\bvolume{4}
\bpages{176--190}.
\end{barticle}
\bptok{imsref}%
\endbibitem

\bibitem{bowsher2007modelling}
\begin{barticle}[mr]
\bauthor{\bsnm{Bowsher},~\bfnm{Clive~G.}\binits{C.~G.}}
(\byear{2007}).
\btitle{Modelling security market events in continuous time: Intensity based, multivariate point process models}.
\bjournal{J. Econometrics}
\bvolume{141}
\bpages{876--912}.
\bid{doi={10.1016/j.jeconom.2006.11.007}, issn={0304-4076}, mr={2413490}}
\end{barticle}
\bptok{imsref}%
\endbibitem

\bibitem{bremaud2001hawkes}
\begin{barticle}[mr]
\bauthor{\bsnm{Br{\'e}maud},~\bfnm{Pierre}\binits{P.}} \AND
\bauthor{\bsnm{Massouli{\'e}},~\bfnm{Laurent}\binits{L.}}
(\byear{2001}).
\btitle{Hawkes branching point processes without ancestors}.
\bjournal{J. Appl. Probab.}
\bvolume{38}
\bpages{122--135}.
\bid{issn={0021-9002}, mr={1816118}}
\end{barticle}
\bptok{imsref}%
\endbibitem

\bibitem{chavez2005point}
\begin{barticle}[mr]
\bauthor{\bsnm{Chavez-Demoulin},~\bfnm{V.}\binits{V.}},
\bauthor{\bsnm{Davison},~\bfnm{A.~C.}\binits{A.~C.}} \AND
\bauthor{\bsnm{McNeil},~\bfnm{A.~J.}\binits{A.~J.}}
(\byear{2005}).
\btitle{Estimating value-at-risk: A point process approach}.
\bjournal{Quant. Finance}
\bvolume{5}
\bpages{227--234}.
\bid{doi={10.1080/14697680500039613}, issn={1469-7688}, mr={2240248}}
\end{barticle}
\bptok{imsref}%
\endbibitem

\bibitem{cox1985theory}
\begin{bmisc}[auto:STB|2014/02/12|14:17:21]
\bauthor{\bsnm{Cox},~\bfnm{J.~C.}\binits{J.~C.}},
\bauthor{\bsnm{IngersollJr},~\bfnm{J.~E.}\binits{J.~E.}} \AND
\bauthor{\bsnm{Ross},~\bfnm{S.~A.}\binits{S.~A.}}
(\byear{1985}).
\bhowpublished{A theory of the term structure of interest rates.
\textit{Econometrica} \textbf{53} 385--407}.
\end{bmisc}
\bptok{imsref}%
\endbibitem

\bibitem{daley2007introduction}
\begin{bbook}[mr]
\bauthor{\bsnm{Daley},~\bfnm{D.~J.}\binits{D.~J.}} \AND
\bauthor{\bsnm{Vere-Jones},~\bfnm{D.}\binits{D.}}
(\byear{1988}).
\btitle{An Introduction to the Theory of Point Processes}.
\bpublisher{Springer},
\blocation{New York}.
\bid{mr={0950166}}
\bptnote{check year}%
\end{bbook}
\bptok{imsref}%
\endbibitem

\bibitem{embrechts2011multivariate}
\begin{barticle}[mr]
\bauthor{\bsnm{Embrechts},~\bfnm{Paul}\binits{P.}},
\bauthor{\bsnm{Liniger},~\bfnm{Thomas}\binits{T.}} \AND
\bauthor{\bsnm{Lin},~\bfnm{Lu}\binits{L.}}
(\byear{2011}).
\btitle{Multivariate {H}awkes processes: An application to financial data}.
\bjournal{J. Appl. Probab.}
\bvolume{48A}
\bpages{367--378}.
\bid{doi={10.1239/jap/1318940477}, issn={0021-9002}, mr={2865638}}
\end{barticle}
\bptok{imsref}%
\endbibitem

\bibitem{errais2010affine}
\begin{barticle}[mr]
\bauthor{\bsnm{Errais},~\bfnm{Eymen}\binits{E.}},
\bauthor{\bsnm{Giesecke},~\bfnm{Kay}\binits{K.}} \AND
\bauthor{\bsnm{Goldberg},~\bfnm{Lisa~R.}\binits{L.~R.}}
(\byear{2010}).
\btitle{Affine point processes and portfolio credit risk}.
\bjournal{SIAM J. Financial Math.}
\bvolume{1}
\bpages{642--665}.
\bid{doi={10.1137/090771272}, issn={1945-497X}, mr={2719785}}
\end{barticle}
\bptok{imsref}%
\endbibitem

\bibitem{filimonov2012quantifying}
\begin{barticle}[auto:STB|2014/02/12|14:17:21]
\bauthor{\bsnm{Filimonov},~\bfnm{V.}\binits{V.}} \AND
\bauthor{\bsnm{Sornette},~\bfnm{D.}\binits{D.}}
(\byear{2012}).
\btitle{Quantifying reflexivity in financial markets: Toward a prediction of flash crashes}.
\bjournal{Phys. Rev. E (3)}
\bvolume{85}
\bpages{056108}.
\end{barticle}
\bptok{imsref}%
\endbibitem

\bibitem{filimonov2013apparent}
\begin{bmisc}[mr]
\bauthor{\bsnm{Filimonov},~\bfnm{V.}\binits{V.}} \AND
\bauthor{\bsnm{Sornette},~\bfnm{D.}\binits{D.}}
(\byear{2013}).
\bhowpublished{Apparent criticality and calibration issues in the {H}awkes
self-excited point process model: Application to high-frequency financial
data. Preprint. Available at \arxivurl{arXiv:1308.6756}.}
\end{bmisc}
\bptok{imsref}%
\endbibitem

\bibitem{hardiman2013critical}
\begin{bmisc}[auto:STB|2014/02/12|14:17:21]
\bauthor{\bsnm{Hardiman},~\bfnm{S.~J.}\binits{S.~J.}},
\bauthor{\bsnm{Bercot},~\bfnm{N.}\binits{N.}} \AND
\bauthor{\bsnm{Bouchaud},~\bfnm{J.-P.}\binits{J.-P.}}
(\byear{2013}).
\bhowpublished{Critical reflexivity in financial markets: A Hawkes process analysis.
Preprint. Available at \arxivurl{arXiv:1302.1405}}.
\end{bmisc}
\bptok{imsref}%
\endbibitem

\bibitem{hawkes1971point}
\begin{barticle}[mr]
\bauthor{\bsnm{Hawkes},~\bfnm{Alan~G.}\binits{A.~G.}}
(\byear{1971}).
\btitle{Point spectra of some mutually exciting point processes}.
\bjournal{J. R. Stat. Soc. Ser. B Stat. Methodol.}
\bvolume{33}
\bpages{438--443}.
\bid{issn={0035-9246}, mr={0358976}}
\end{barticle}
\bptok{imsref}%
\endbibitem

\bibitem{hawkes1971spectra}
\begin{barticle}[mr]
\bauthor{\bsnm{Hawkes},~\bfnm{Alan~G.}\binits{A.~G.}}
(\byear{1971}).
\btitle{Spectra of some self-exciting and mutually exciting point processes.}
\bjournal{Biometrika}
\bvolume{58}
\bpages{83--90}.
\bid{issn={0006-3444}, mr={0278410}}
\end{barticle}
\bptok{imsref}%
\endbibitem

\bibitem{hawkes1974cluster}
\begin{barticle}[mr]
\bauthor{\bsnm{Hawkes},~\bfnm{Alan~G.}\binits{A.~G.}} \AND
\bauthor{\bsnm{Oakes},~\bfnm{David}\binits{D.}}
(\byear{1974}).
\btitle{A cluster process representation of a self-exciting process}.
\bjournal{J. Appl. Probab.}
\bvolume{11}
\bpages{493--503}.
\bid{issn={0021-9002}, mr={0378093}}
\end{barticle}
\bptok{imsref}%
\endbibitem

\bibitem{heston1993closed}
\begin{barticle}[auto:STB|2014/02/12|14:17:21]
\bauthor{\bsnm{Heston},~\bfnm{S.~L.}\binits{S.~L.}}
(\byear{1993}).
\btitle{A closed-form solution for options with stochastic volatility with applications to bond and currency options}.
\bjournal{Review of Financial Studies}
\bvolume{6}
\bpages{327--343}.
\end{barticle}
\bptok{imsref}%
\endbibitem

\bibitem{hewlett2006clustering}
\begin{bincollection}[auto:STB|2014/02/12|14:17:21]
\bauthor{\bsnm{Hewlett},~\bfnm{P.}\binits{P.}}
(\byear{2006}).
\btitle{Clustering of order arrivals, price impact and trade path optimisation}.
In \bbooktitle{Workshop on Financial Modeling with Jump Processes,
6--8 September 2006}.
\bpublisher{Ecole Polytechnique},
\blocation{France}.
\end{bincollection}
\bptok{imsref}%
\endbibitem

\bibitem{jacod1975multivariate}
\begin{barticle}[mr]
\bauthor{\bsnm{Jacod},~\bfnm{Jean}\binits{J.}}
(\byear{1974/75}).
\btitle{Multivariate point processes: Predictable projection, {R}adon--{N}ikod\'ym derivatives, representation of martingales}.
\bjournal{Z. Wahrsch. Verw. Gebiete}
\bvolume{31}
\bpages{235--253}.
\bid{mr={0380978}}
\bptnote{check year}%
\end{barticle}
\bptok{imsref}%
\endbibitem

\bibitem{jacod1987limit}
\begin{bbook}[mr]
\bauthor{\bsnm{Jacod},~\bfnm{Jean}\binits{J.}} \AND
\bauthor{\bsnm{Shiryaev},~\bfnm{Albert~N.}\binits{A.~N.}}
(\byear{1987}).
\btitle{Limit Theorems for Stochastic Processes}.
\bseries{Grundlehren der Mathematischen Wissenschaften}
\bvolume{288}.
\bpublisher{Springer},
\blocation{Berlin}.
\bid{mr={0959133}}
\end{bbook}
\bptok{imsref}%
\endbibitem

\bibitem{jaisson2013impact}
\begin{bmisc}[auto:STB|2014/02/12|14:17:21]
\bauthor{\bsnm{Jaisson},~\bfnm{T.}\binits{T.}}
(\byear{2013}).
\bhowpublished{Market impact as anticipation of the order flow imbalance. Working paper}.
\end{bmisc}
\bptok{imsref}%
\endbibitem

\bibitem{appendix}
\begin{bmisc}[auto:STB|2014/02/12|14:17:21]
\bauthor{\bsnm{Jaisson},~\bfnm{T.}\binits{T.}} \AND
\bauthor{\bsnm{Rosenbaum},~\bfnm{M.}\binits{M.}}
(\byear{2014}).
\bhowpublished{Limit theorems for nearly unstable Hawkes processes: Version with technical appendix.
Technical Report 1607, Laboratoire de Probabilit\'es et Mod\`eles Al\'eatoires, Univ. Pierre et Marie Curie}.
\end{bmisc}
\bptok{imsref}%
\endbibitem

\bibitem{jakubowski1989convergence}
\begin{barticle}[mr]
\bauthor{\bsnm{Jakubowski},~\bfnm{A.}\binits{A.}},
\bauthor{\bsnm{M{\'e}min},~\bfnm{J.}\binits{J.}} \AND
\bauthor{\bsnm{Pag{\`e}s},~\bfnm{G.}\binits{G.}}
(\byear{1989}).
\btitle{Convergence en loi des suites d'int\'egrales stochastiques sur
l'espace {$\mathbf{D}^1$} de {S}korokhod}.
\bjournal{Probab. Theory Related Fields}
\bvolume{81}
\bpages{111--137}.
\bid{doi={10.1007/BF00343739}, issn={0178-8051}, mr={0981569}}
\end{barticle}
\bptok{imsref}%
\endbibitem

\bibitem{kalashnikov1997geometric}
\begin{bbook}[mr]
\bauthor{\bsnm{Kalashnikov},~\bfnm{Vladimir}\binits{V.}}
(\byear{1997}).
\btitle{Geometric Sums: Bounds for Rare Events with Applications: Risk Analysis, Reliability, Queueing}.
\bseries{Mathematics and Its Applications}
\bvolume{413}.
\bpublisher{Kluwer},
\blocation{Dordrecht}.
\bid{mr={1471479}}
\end{bbook}
\bptok{imsref}%
\endbibitem

\bibitem{kurtz1991weak}
\begin{barticle}[mr]
\bauthor{\bsnm{Kurtz},~\bfnm{Thomas~G.}\binits{T.~G.}} \AND
\bauthor{\bsnm{Protter},~\bfnm{Philip}\binits{P.}}
(\byear{1991}).
\btitle{Weak limit theorems for stochastic integrals and stochastic differential equations}.
\bjournal{Ann. Probab.}
\bvolume{19}
\bpages{1035--1070}.
\bid{issn={0091-1798}, mr={1112406}}
\end{barticle}
\bptok{imsref}%
\endbibitem

\bibitem{large2007measuring}
\begin{barticle}[auto:STB|2014/02/12|14:17:21]
\bauthor{\bsnm{Large},~\bfnm{J.}\binits{J.}}
(\byear{2007}).
\btitle{Measuring the resiliency of an electronic limit order book}.
\bjournal{Journal of Financial Markets}
\bvolume{10}
\bpages{1--25}.
\end{barticle}
\bptok{imsref}%
\endbibitem

\bibitem{lillo2004long}
\begin{barticle}[auto:STB|2014/02/12|14:17:21]
\bauthor{\bsnm{Lillo},~\bfnm{F.}\binits{F.}} \AND
\bauthor{\bsnm{Farmer},~\bfnm{J.~D.}\binits{J.~D.}}
(\byear{2004}).
\btitle{The long memory of the efficient market}.
\bjournal{Stud. Nonlinear Dyn. Econom.}
\bvolume{8}
\bpages{3}.
\end{barticle}
\bptok{imsref}%
\endbibitem

\bibitem{ogata1978asymptotic}
\begin{barticle}[mr]
\bauthor{\bsnm{Ogata},~\bfnm{Yosihiko}\binits{Y.}}
(\byear{1978}).
\btitle{The asymptotic behaviour of maximum likelihood estimators for stationary point processes}.
\bjournal{Ann. Inst. Statist. Math.}
\bvolume{30}
\bpages{243--261}.
\bid{doi={10.1007/BF02480216}, issn={0020-3157}, mr={0514494}}
\end{barticle}
\bptok{imsref}%
\endbibitem

\bibitem{ogata1983likelihood}
\begin{barticle}[mr]
\bauthor{\bsnm{Ogata},~\bfnm{Yoshihiko}\binits{Y.}}
(\byear{1983}).
\btitle{Likelihood analysis of point processes and its applications to seismological data}.
\bjournal{Bull. Inst. Internat. Statist.}
\bvolume{}
\bvolume{50}
\bpages{943--961}.
\end{barticle}
\bptok{imsref}%
\endbibitem

\bibitem{petrov1975sums}
\begin{bbook}[mr]
\bauthor{\bsnm{Petrov},~\bfnm{V.~V.}\binits{V.~V.}}
(\byear{1975}).
\btitle{Sums of Independent Random Variables}.
\bpublisher{Springer},
\blocation{New York}.
\bid{mr={0388499}}
\end{bbook}
\bptok{imsref}%
\endbibitem

\bibitem{reynaud2010adaptive}
\begin{barticle}[mr]
\bauthor{\bsnm{Reynaud-Bouret},~\bfnm{Patricia}\binits{P.}} \AND
\bauthor{\bsnm{Schbath},~\bfnm{Sophie}\binits{S.}}
(\byear{2010}).
\btitle{Adaptive estimation for {H}awkes processes; application to genome analysis}.
\bjournal{Ann. Statist.}
\bvolume{38}
\bpages{2781--2822}.
\bid{doi={10.1214/10-AOS806}, issn={0090-5364}, mr={2722456}}
\end{barticle}
\bptok{imsref}%
\endbibitem

\bibitem{shah2013results}
\begin{barticle}[mr]
\bauthor{\bsnm{Shah},~\bfnm{P.~V.}\binits{P.~V.}} \AND
\bauthor{\bsnm{Jana},~\bfnm{R.~K.}\binits{R.~K.}}
(\byear{2013}).
\btitle{Results on generalized {M}ittag--{L}effler function via {L}aplace transform}.
\bjournal{Appl. Math. Sci. (Ruse)}
\bvolume{7}
\bpages{567--570}.
\bid{issn={1312-885X}, mr={3007505}}
\end{barticle}
\bptok{imsref}%
\endbibitem

\bibitem{wyart2008relation}
\begin{barticle}[auto:STB|2014/02/12|14:17:21]
\bauthor{\bsnm{Wyart},~\bfnm{M.}\binits{M.}},
\bauthor{\bsnm{Bouchaud},~\bfnm{J.-P.}\binits{J.-P.}},
\bauthor{\bsnm{Kockelkoren},~\bfnm{J.}\binits{J.}},
\bauthor{\bsnm{Potters},~\bfnm{M.}\binits{M.}} \AND
\bauthor{\bsnm{Vettorazzo},~\bfnm{M.}\binits{M.}}
(\byear{2008}).
\btitle{Relation between bid--ask spread, impact and volatility in order-driven markets}.
\bjournal{Quant. Finance}
\bvolume{8}
\bpages{41--57}.
\end{barticle}
\bptok{imsref}%
\endbibitem


\bibitem{zhu2013central}
\begin{barticle}[auto]
\bauthor{\bsnm{Zhu},~\bfnm{Lingjiong}\binits{L.}}
(\byear{2013}).
\btitle{Central limit theorem for nonlinear Hawkes processes}.
\bjournal{J. Appl. Probab.}
\bvolume{50}
\bpages{760--771}.
\bid{mr={3102513}}
\end{barticle}
\bptok{imsref}%
\endbibitem

\end{thebibliography}
\end{document}